\documentclass[aps,twocolumn,showpacs,groupedaddress,nofootinbib]{revtex4}
\usepackage{epsfig}
\usepackage{bm}
\usepackage{amsmath}
\usepackage{wasysym}

\newcommand{\ds}{\displaystyle}

\newcommand{\be}{\begin{equation}}
\newcommand{\ee}{\end{equation}}
\newcommand{\ba}{\begin{eqnarray}}
\newcommand{\ea}{\end{eqnarray}}
\newcommand{\etal}{\mbox{\it et al.}}

\newcommand{\rp}{\textsc{r/p}-}

\begin{document}

\preprint{02-02}

\title{
Deep exclusive charged $\pi$ electroproduction above the resonance region}
\author{Murat M. Kaskulov}
\email{kaskulov@theo.physik.uni-giessen.de}
\author{Ulrich Mosel}
\affiliation{Institut f\"ur Theoretische Physik, Universit\"at Giessen,
             D-35392 Giessen, Germany}
\date{\today}

\begin{abstract}
A description of exclusive charged pion electroproduction  $(e,e'\pi^{\pm})$
off nucleons at high energies is proposed. The  model combines a Regge pole 
approach with residual effect of nucleon resonances. The exchanges of 
$\pi$(140), vector $\rho(770)$ and axial-vector $a_1(1260)$ and $b_1(1235)$ 
Regge trajectories are considered. The contribution of nucleon resonances 
is described using a dual connection between the exclusive hadronic form 
factors and inclusive deep inelastic structure functions. The model describes 
the measured longitudinal, transverse and interference cross sections at JLAB 
and DESY. The scaling behavior of the cross sections is in agreement with JLAB 
and deeply virtual HERMES data. The results for a polarized beam-spin
azimuthal asymmetry in $(\vec{e},e'\pi^{\pm})$ are presented. Model
predictions for JLAB at 12 GeV are given. 
\end{abstract}
\maketitle


\section{Introduction}

At JLAB the exclusive reaction $p(e,e'\pi^+)n$  has been studied for a 
wide range of photon virtualities $Q^2$ at an invariant mass of the $\pi^+n$
system around the onset of deep inelastic scattering (DIS) regime,
$W\simeq 2$~GeV~\cite{Horn:2006tm,Tadevosyan:2007yd,Horn:2007ug,Blok:2008jy,Huber:2008id}.
A separation of the cross section into the transverse $\sigma_{\rm T}$,
longitudinal $\sigma_{\rm L}$ and interference $\sigma_{\rm TT}$ and
$\sigma_{\rm LT}$ components has been performed. The CLAS data for the polarized
beam single-spin asymmetry in $p(\vec{e},e'\pi^+)n$ are also 
available~\cite{Avakian:2004dt}. The HERMES data at DESY~\cite{:2007an} 
extend the kinematic region to much higher values of $W^2>$10~GeV$^2$ toward 
the true DIS region $Q^2\gg 1$~GeV$^2$ and much higher values of $-t$. The
cross section for $p(e,e'\pi^+)n$ has also been measured above the resonance region
at the Cambridge Electron Accelerator (CEA)~\cite{CEA}, in $p(e,e'\pi^+)n$ and 
$n(e,e'\pi^-)p$ at the Wilson Synchrotron Laboratory at 
Cornell~\cite{Cornell_1,Cornell_2,Cornell_3} and 
DESY~\cite{Ackermann:1977rp,DESY1,DESY2,DESY3,DESY4}.

The longitudinal cross section $\sigma_{\rm L}$ is generally thought to be
well understood in terms of the pion quasi-elastic knockout 
mechanism~\cite{Neudatchin:2004pu} because of the pion pole at low $-t$ . 
If true, this makes it possible to study the charge form factor of the pion 
at momentum transfer much bigger than in the scattering of pions from atomic 
electrons~\cite{Sullivan:1970yq}. On the other hand, the transverse cross section
$\sigma_{\rm T}$ is predicted to be suppressed by $\sim 1/Q^2$ with respect to 
$\sigma_{\rm L}$ for sufficiently high values of $Q^2$ and $W$~\cite{Collins:1996fb}.
On the experimental side, however, the JLAB data show that at forward angles
$\sigma_{\rm T}$ is large. For instance, at $Q^2 =
3.91$~GeV$^2$~\cite{Horn:2007ug} $\sigma_{\rm T}$ is by about a factor of two
larger than $\sigma_{\rm L}$ and at $Q^2=2.15$~GeV$^2$ it has same size as 
$\sigma_{\rm L}$ in agreement with previous JLAB measurements~\cite{Horn:2006tm}.

There is a long standing issue concerning the reaction mechanisms contributing
to deeply virtual $\pi$ electroproduction  above the resonance 
region~\cite{Nachtmann:1976be,Collins:1980dv,Faessler:2007bc}. The models which describe 
$(e,e'\pi^{\pm})$  in terms of hadronic degrees of freedom fail to reproduce 
$\sigma_{\rm T}$ observed in these reactions, see Ref.~\cite{Blok:2008jy} and 
references therein. Previous measurements~\cite{Horn:2006tm,Tadevosyan:2007yd,Cornell_3}
at smaller and much higher values of $Q^2$ show a similar problem in the 
understanding of $\sigma_{\rm T}$. Already from values of $Q^2>0.6$~GeV$^2$ 
the meson-exchange and/or Regge pole models are not compatible with the
measured interference $\sigma_{\rm TT}$ and $\sigma_{\rm LT}$ cross sections
and the extraction of the pion form factor relies on the fit to the
longitudinal cross section $\sigma_{\rm L}$ only~\cite{Huber:2008id}. A
remarkably rich experimental data base obtained for $N(e,e'\pi)N'$ above the 
resonance region remains unexplained~\cite{Horn:2007ug,Blok:2008jy}. On the
other hand, a detailed knowledge of the $p(e,e'\pi^+)n$ reaction above the 
resonances $\sqrt{s}>2$~GeV is mandatory for the interpretation of the color 
transparency signal observed in this reaction off 
nuclei~\cite{:2007gqa,Kaskulov:2008ej}.

A possible description of $\sigma_{\rm T}$ at JLAB has been proposed in 
Ref.~\cite{Kaskulov:2008xc}. The approach followed there is to complement 
the hadron-like interaction types in the $t$-channel, which dominate in 
photoproduction and low $Q^2$ electroproduction, with the direct interaction 
of virtual photons with partons followed by string (quark) fragmentation 
into $\pi^+n$. Then $\sigma_{\rm T}$ can be readily explained and both 
$\sigma_{\rm L}$ and $\sigma_{\rm T}$ can be described from low up to high 
values of $Q^2$. In~\cite{Kaskulov:2008xc} the reaction $p(e,e'\pi^+)n$ is 
treated as an exclusive limit, $z \to 1$, of semi-inclusive DIS 
\be
p(e,e'\pi^+)X \overset{z\to 1}{\longrightarrow}  p(e,e'\pi^+)n
\ee 
in the spirit of an exclusive-inclusive connection~\cite{Bjorken:1973gc}. 
The transverse cross section in $n(e,e',\pi^-)p$ has been predicted to be smaller than
in $p(e,e',\pi^+)n$.  The model~\cite{Kaskulov:2009gp} has also been applied
to values of $(Q^2,W)$ in the DIS region at HERMES~\cite{:2007an}. 
In~\cite{Kaskulov:2009gp} $\sigma_{\rm T}$ in DIS  gets much smaller in the 
forward $\pi^+$ production, but still dominates the off-forward region. 

However, in~\cite{Kaskulov:2008xc,Kaskulov:2009gp} the transverse cross
section $\sigma_{\rm T}$ itself was modeled and the solution of the problem 
on the amplitude level is still missing.  Both the soft hadronic and hard 
partonic parts of the amplitude can in principle interfere making non-additive 
contributions to $\sigma_{\rm L}$ and to interference $\sigma_{\rm TT}$ and 
$\sigma_{\rm LT}$ cross sections. One might describe this transverse strength 
in the language of perturbative QCD by considering higher twist corrections 
to a generalized parton distribution (GPD) based handbag diagram.
This approach has been followed in Ref.~\cite{Goloskokov:2009ia}
where $p(\gamma^*,\pi^+)n$ is considered using the handbag approach
with a $\pi$-pole contribution. Indeed, the data from JLAB 
demonstrate~\cite{Horn:2007ug,Blok:2008jy} that the magnitude and sign of the
interference cross sections are not compatible with the simple exchange of a 
pion trajectory in the $t$-channel. Because, the contributions from exchange
of heavy mesons are small~\cite{Kaskulov:2008xc} this would suggest the
presence of a large transverse resonance or partonic  interfering background 
to the meson-pole contributions.

In this work we attempt a phenomenological approach to model the hard
scattering or, using a duality argument, the presence of nucleon resonances 
beyond the $t$-channel meson-pole amplitudes. The meson-exchange processes 
dominate in high-energy photoproduction and low $Q^2$ electroproduction 
above the resonance region. One way to describe this region is to assume 
that the coherent sum of baryon resonance contributions would be expected 
by duality arguments to be equivalent to a sum over $t$-channel Regge 
trajectories. However, in electroproduction, with plausible assumptions 
concerning the coupling constants and transition form factors, the exchange 
of heavy mesons alone does not explain the transverse cross section and turns out 
to be marginal~\cite{Kaskulov:2008xc}. It is also a generic rule that single 
$t$-channel meson-exchange processes vanish in the forward $\pi^+$ direction.
On the other hand, pion exchange does play an important role at near forward 
directions and must be included, as must be the nucleon-pole charge term to 
satisfy gauge invariance. The nucleon magnetic transitions vanish 
in the forward production and can be neglected~\cite{Guidal:1997hy}. For
instance, in photoproduction this suggests an extreme phenomenological
scenario, known as an electric model, where the only relevant contributions 
to $\pi^{\pm}$ production at forward angles are the ones from $\pi$-exchange and 
the nucleon Born term where the inclusion of the latter is mandatory to
conserve gauge invariance.

By reggeization of the $\pi$-exchange one takes into account higher mass and
higher spin excitations. At forward angles considered here the momentum
transfer $-t$ is small and the exchanged $\pi$-trajectory is close to its 
first materialization. However, the nucleons in the $s(u)$-channel pole  
amplitudes are highly off-mass-shell and with increasing values of $(Q^2,W)$ 
the effect of nucleon resonances should become more and more important.
This is because of the well known hardening of the higher mass 
resonance transition form factors which must respect the scaling properties of
deep inelastic structure functions in inclusive  
scattering~\cite{Bloom:1970xb,Bloom:1971ye}. We shall follow this suggestion and 
model the contribution of nucleon resonances using a local Bloom-Gilman
connection between the exclusive and inclusive processes.

Another question which we address here is a possible contribution of the 
resonance (or partonic) background to the longitudinal cross 
section $\sigma_{\rm L}$ which is presently used to get the information 
about the pion form factor. 
Indeed, the same {\it resonance/partonic} background also affects the 
longitudinal cross section making the $\pi$-pole  dominance
in the longitudinal response rather questionable. 
Based on a quantitative description of
electroproduction data achieved in this work in a large range of $(Q^2,W)$ 
from JLAB to DIS region at HERMES the present results may assist in the
experimental  analysis  and extraction of the pion charge form factor to 
minimize systematic uncertainties. Recall that it is essential to use 
theoretical model input for the extraction of the pion form factor~\cite{Huber:2008id}.

The outline of the present paper is as follows. In the Section II  we 
briefly recall the kinematics and definition of the cross sections in 
exclusive $(e,e'\pi^{\pm})$ electroproduction reaction. In  Section 
III we discuss our treatment of gauge invariant $\pi$-exchange in the 
Regge pole model. In Section IV we consider the effect of nucleon 
resonances and derive the transition form factors using an 
exclusive-inclusive connection. In Section V we consider the 
contribution of vector $\rho(770)$ and axial-vector $a_1(1260)$   and 
$b_1(1235)$ Regge trajectories. In Section VI we briefly discuss  the 
$\pi^{\pm}$ photoproduction at forward angles. Then the model is 
extended to electroproduction. In Sections VII-IX the results are 
compared to the experimental data from JLAB, DESY and Cornell. In 
Section X we compare our results with the HERMES data. The $Q^2$ behavior 
of the cross sections is studied in Section XI. The polarized beam-spin
asymmetry and the role played  by  the axial-vector mesons in $(\vec{e},e'\pi^{\pm})$ 
are discussed in Section XII.  In Section XIII the model predictions 
for JLAB at 12 GeV are presented. The conclusions are summarized in 
Section XIV. Some details of the calculations are relegated to the Appendix.

\begin{figure}[t]
\begin{center}
\includegraphics[clip=true,width=0.95\columnwidth,angle=0.]{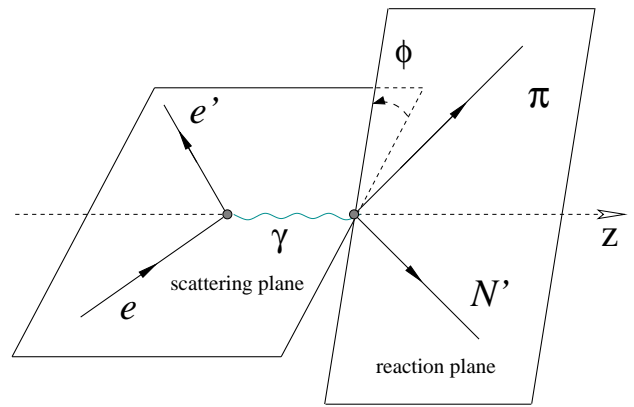}
\caption{\label{eepin} 
Exclusive reaction $N(e,e'\pi)N'$ in the laboratory. $\phi$ stands for the 
azimuthal angle between the  electron scattering $(e,e')$ plane and reaction 
$N(\gamma^*,\pi)N'$ plane.}
\vspace{-0.5cm}
\end{center}
\end{figure}

\section{Kinematics and definitions}

We recall briefly the kinematics in exclusive $\pi$ electroproduction
\begin{equation}
\label{eepireac}
e(l) + N(p) \to e'(l') + \pi(k') + N'(p'),
\end{equation}
and specify the notations and definitions of variables. The reaction 
(\ref{eepireac}) in the laboratory is shown in Figure~\ref{eepin} where 
the target nucleon is at rest, the $z$-axis is directed along the three 
momentum $\vec{q}=(0,0,\sqrt{\nu^2+Q^2})$ of the exchanged virtual photon 
$\gamma^*$ with $q= l-l'= (\nu,\vec{q})$, $Q^2=-q^2$, $\nu=E_e-E_e'$  and
$l(l')$ is the four momentum of incoming (deflected) electrons. In 
Figure~\ref{eepin} $\phi$ stands for the azimuthal angle between the 
electron scattering $(e,e')$ plane and $\gamma^* N \to \pi N'$ reaction 
plane. $\phi$ is zero when the pion is closest to the outgoing  
electron~\cite{Bacchetta:2004jz}.

In exclusive reaction $(e,e'\pi)$ we shall deal with an unpolarized target 
and, both unpolarized and polarized lepton beams. The differential cross 
section is given by
\begin{eqnarray}
\label{dsdte}
\frac{d\sigma}{dQ^2d\nu dt d\phi} &=&
\frac{\Phi}{2\pi} 
\left[
           \frac{d\sigma_{\rm T}}{dt}
+ \varepsilon \frac{d\sigma_{\rm L}}{dt} \right. \nonumber \\
&+& \left.  \sqrt{2\varepsilon (1+\varepsilon)}
\frac{d\sigma_{\rm LT}}{dt} \cos(\phi)  \right. \nonumber \\
&+& \left.
\varepsilon \frac{d\sigma_{\rm TT}}{dt} \cos(2\phi)
  \right. \nonumber \\
&+& \left.  h \sqrt{2\varepsilon(1-\varepsilon)}
\frac{d\sigma_{\rm LT'}}{dt} \sin(\phi)
\right],
\end{eqnarray}
where $d\sigma_{\rm T}$ is the transverse cross section, $d\sigma_{\rm L}$ is 
the longitudinal cross section, $d\sigma_{\rm TT}$ is the cross section 
originating from the interference between the transverse components of the 
virtual photon, $d\sigma_{\rm LT}$ is the cross section arising from the 
interference between the transverse and longitudinal polarizations of the 
virtual photon and $d\sigma_{\rm LT'}$ is the beam-spin polarized cross section
resulting from the interference between the transverse and longitudinal
photons and helicity $h=\pm 1$  of the incoming electron.

The virtual photon flux is conventionally defined as~\cite{Hand:1963bb}
\begin{equation}
\Phi = \frac{\pi}{E_e(E_e-\nu)}
\left(\frac{\alpha_e}{2\pi^2} \frac{E_e-\nu}{E_e} \frac{\mathcal K}{Q^2}
\frac{1}{1-\varepsilon}\right),
\end{equation}
with $\alpha_e \simeq 1/137$, $\mathcal K = (W^2-M^2_N)/2M_N$  and
\begin{equation}
\varepsilon = \frac{1}{ 1+2 \frac{\nu^2+Q^2}{4E_e(E_e-\nu)-Q^2}}
\end{equation}
is the ratio of longitudinal to transverse polarization of the virtual photon. 
The longitudinal/transverse (\textsc{l/t}) separated virtual-photon nucleon 
cross sections are given in Appendix~\ref{appSCS}. The
$t$-differential cross section for $N(\gamma^*,\pi)N'$ integrated over $\phi$ is 
denoted here as
\be
\label{CSU}
\frac{d\sigma_{\rm U}}{dt}=\frac{d\sigma_{\rm T}}{dt}+\varepsilon \frac{d\sigma_{\rm L}}{dt}.
\ee

\begin{figure*}[t]
\begin{center}
\includegraphics[clip=true,width=2.05\columnwidth,angle=0.]{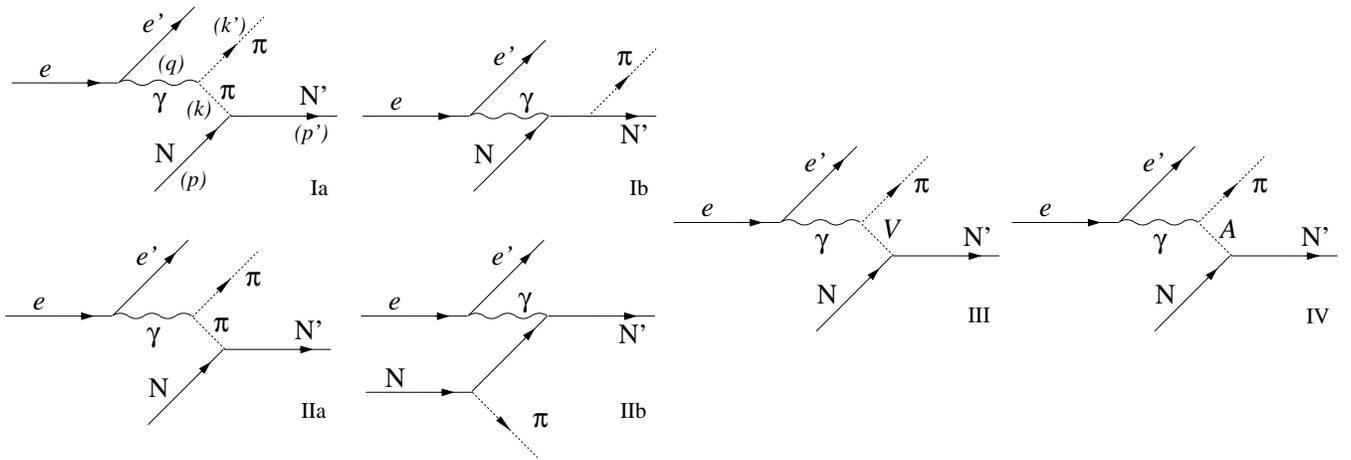}
\caption{\label{DiagramsEffRegge} 
\small The diagrams describing the $\pi^+$ and/or $\pi^-$ electroproduction amplitudes in
exclusive reactions $p(e,e'\pi^+)n$ and $n(e,e'\pi^-)p$.
In $p(e,e'\pi^+)n$ the $s$-channel nucleon-pole term (\textsc{ib} diagram) is added to
the $t$-channel $\pi$-pole exchange (\textsc{ia} diagram) to conserve the charge of
the system. Similarly, the diagram \textsc{iia} and the $u$-channel
nucleon-pole diagram \textsc{iib} form gauge invariant amplitude in
$n(e,e'\pi^-)p$. The last two diagrams (\textsc{iii} and \textsc{iv})
correspond to the exchange of vector $V=\rho(770)$ and axial-vector $A =
a_1(1260)$ and $b_{1}(1235)$  Regge trajectories. The momentum flows are shown
in the diagram \textsc{ia}.
\vspace{-0.2cm}
}
\end{center}
\end{figure*}

The longitudinal beam single-spin asymmetry (SSA) in $(\vec{e},e'\pi)$ scattering is defined
so that
\be
\label{BSSA}
A_{\rm LU}(\phi) \equiv
\frac{d\sigma^{\rightarrow}(\phi)-d\sigma^{\leftarrow}(\phi)}{d\sigma^{\rightarrow}(\phi)+d\sigma^{\leftarrow}(\phi)},
\ee
where $d\sigma^{\rightarrow}$ refers to positive helicity $h=+1$ of the incoming
electron. The azimuthal moment associated with the beam SSA is
given by~\cite{Bacchetta:2004jz}
\be
\label{BeamSSAmoment}
A^{\sin(\phi)}_{\rm LU} = \frac{\sqrt{2\varepsilon(1-\varepsilon)}d\sigma_{\rm
    LT'}}{d\sigma_{\rm T} + \varepsilon d\sigma_{\rm L}}.
\ee

\section{Gauging the pion exchange}

The diagrams describing the $\pi^+$ and/or $\pi^-$ electroproduction amplitudes in
exclusive reactions $p(e,e'\pi^+)n$ and $n(e,e'\pi^-)p$ are shown in 
Figure~\ref{DiagramsEffRegge}. At high energies the particles exchanged
in the $t$-channel are understood as the Regge trajectories. In
$p(e,e'\pi^+)n$ the $s$-channel nucleon-pole term (\textsc{ib} diagram) is added to
the $t$-channel $\pi$-pole exchange (\textsc{ia} diagram) to conserve the charge of
the system. Similarly, the diagram \textsc{iia}  and the $u$-channel nucleon-pole
diagram \textsc{iib}  form a gauge invariant amplitude in $n(e,e'\pi^-)p$. The
last two diagrams (\textsc{iii} and \textsc{iv}) correspond to the exchange of 
vector $V=\rho(770)$ and axial-vector $A = a_1(1260)$, $b_{1}(1235)$  Regge 
trajectories.

The $\pi$-exchange currents describing the reactions $p(\gamma^*,\pi^+)n$ and 
$n(\gamma^*,\pi^-)p$ take the form~\cite{Kaskulov:2008ej}
\begin{widetext}
\ba
\label{PipolePl}
{-iJ^{\mu}_s}(\gamma^*p\to \pi^+ n) =  {\sqrt{2} g_{\pi NN}}\,
\bar{u}_{s'}(p')\gamma_5
\left[
\mathcal{F}_{\gamma\pi\pi}(Q^2,t) \frac{ (k+k')^{\mu}}{t-m_{\pi}^2+i0^+}
\right.
+    \mathcal{F}_{s}(Q^2,s,t)
\frac{(p+q)_{\sigma}\gamma^{\sigma}\gamma^{\mu} + M_p\gamma^{\mu}}
{s-M_p^2+i0^+}
\nonumber \\
+ \left.
[\mathcal{F}_{\gamma\pi\pi}(Q^2,t)-\mathcal{F}_{s}(Q^2,s,t)] \frac{(k-k')^{\mu}}{Q^2}
\right] u_s(p), 
\ea
\ba
\label{PipoleMi}
{-iJ^{\mu}_u}(\gamma^*n\to \pi^- p) =
- {\sqrt{2} g_{\pi NN}} \, \bar{u}_{s'}(p')
\left[
\mathcal{F}_{\gamma\pi\pi}(Q^2,t) \frac{ (k+k')^{\mu}}{t-m_{\pi}^2+i0^+}
\right.
-   \mathcal{F}_{u}(Q^2,u,t)
\frac{ \gamma^{\mu}(p'-q)_{\sigma}\gamma^{\sigma} + M_{p}\gamma^{\mu}
}{u-M_p^2+i0^+}
\nonumber \\
+ \left.
[\mathcal{F}_{\gamma\pi\pi}(Q^2,t)-\mathcal{F}_{u}(Q^2,u,t)] \frac{(k-k')^{\mu}}{Q^2}
\right] \gamma_5 u_s(p),
\ea
where $\mathcal{F}_{\gamma\pi\pi}(Q^2,t)$ denotes the transition form factor
of the pion and $\mathcal{F}_{s(u)}(Q^2,s(u),t)$ stands for the proton $s(u)$-channel
transition form factor. In Eqs.~(\ref{PipolePl}) and (\ref{PipoleMi}) $g_{\pi NN}=13.4$
is the pseudoscalar $\pi N$ coupling constant, $t=k^2$, $s=W^2$, $k=k'-q=p-p'$
and other notations are obvious. 

The amplitudes are gauge invariant and the
current conservation condition, $q_{\mu}J^{\mu}_{s(u)} =0$, is satisfied in the
presence of different form factors, $\mathcal{F}_{\gamma\pi\pi}$ and
$\mathcal{F}_{s(u)}$, which in general can depend on values of
$(Q^2,s(u),t)$. Eqs.~(\ref{PipolePl}) and (\ref{PipoleMi}) are obtained using the
requirement that the modified electromagnetic vertex functions  entering the 
amplitude obey the same Ward-Takahashi identities as the bare 
ones~\cite{Naus:1989em,Koch:2001ii,Gross:1987bu}. Further aspects concerning 
the gauged electric amplitude are relegated to Appendix~\ref{GaugedBorn}.

At high energies the exchange of high-spin and high-mass particles lying on
the $\pi$-Regge trajectory has to be taken into account. Then, to continue 
the electric amplitude to high energies we define the half off-shell form factor
\be
\mathcal{F}_{\gamma\pi\pi}(Q^2,t) = {F}_{\gamma\pi\pi}(Q^2)
(t-m_{\pi}^2) \mathcal{R}(\alpha_{\pi}(t)).
\ee
In the $\pi$-pole term this procedure replaces the Feynman propagator
by the Regge propagator suggested by the high energy limit of the amplitude
\be
\label{ReggeP}
\mathcal{D}(t) = \frac{1}{t-m_{\pi}^2+i0^+} \Longrightarrow
\mathcal{R}(\alpha_{\pi}(t)) = \left[\frac{1 + e^{-i\pi\alpha_{\pi}(t)}}{2}\right]   
\left(- \alpha'_{\pi} \right) \Gamma[-\alpha_{\pi}(t)]
e^{\alpha_{\pi}(t) \ln(\alpha_{\pi}'s)}, 
\ee
where $\alpha_{\pi}(t) = \alpha'_{\pi}(t-m^2_{\pi})$ is the Regge trajectory
of $\pi$ with a slope $\alpha'_{\pi}=0.74$~GeV$^{-2}$ and $\Gamma$ function 
results from suppression of singularities in the physical region. Close to 
the pole position $t\to m^2_{\pi}$  the Regge propagator is reduced to
$1/(t-m_{\pi}^2)$ and we approach the Feynman amplitude describing the first
$\pi(140)$ materialization of the trajectory.

We further treat the nucleon-pole part as an indispensable part of the 
$\pi$-pole amplitude. At the real photon point gauge invariance requires for the 
nucleon-pole term the same phase and $t$-dependence as in 
the $\pi$-Regge amplitude~\cite{Guidal:1997hy}
\be
\label{F1generic}
\mathcal{F}_{s(u)}(Q^2,s(u),t) = {F}_{s(u)}(Q^2,s(u)) (t-m_{\pi}^2) 
 \mathcal{R}(\alpha_{\pi}(t)).
\ee
This assumption  is justified by the observation that there exists a gauge
where the $\pi$-exchange vanishes and the $\pi$-pole contribution is
generated kinematically by the nucleon-pole term itself~\cite{Jones:1979aa}.
\end{widetext}

For the pion transition form factor $F_{\gamma\pi\pi}$ we use a monopole 
parameterization
\begin{equation}
\label{PiFF}
{F}_{\gamma\pi\pi} (Q^2) =
{[1+Q^2/\Lambda_{\gamma\pi\pi}^2]^{-1}},
\end{equation}
with a cut-off $\Lambda_{\gamma\pi\pi}$ as a fit parameter. In general, the
cut-off can be a function of $t$,
$\Lambda_{\gamma\pi\pi}=\Lambda_{\gamma\pi\pi}(t)$, reflecting the
off-shellness of the pion in the $t$-channel and the underlying space-time 
pattern of direct partonic interactions at high values of $-t$~\cite{Laget:2004qu}.
In the forward $\pi^+$ production the momentum transfer $t$ is rather small and
the exchanged pion is close to its mass shell. In the fit to data we shall not
allow large deviations from the VMD value. 

Since the $\pi$-pole contribution is replaced by an exchange of  reggeon-pion, the
relation to the on-shell pion form factor might be lost~\cite{Mankiewicz:1998kg}
 and ${F}_{\gamma\pi\pi} (Q^2)$ should be understood as an effective transition
form factor.

\section{Effect of nucleon resonances}
\label{BGduality}

Similar arguments should apply to the transition form factor in the
$s(u)$-channel nucleon-pole terms. A simplest choice would be to use 
in Eq.~(\ref{F1generic}) 
\begin{equation}
\label{FsF1p}
{F}_{s(u)}(Q^2,s(u)) = {F}_1^p(Q^2), 
\end{equation}
where ${F}_1^p(Q^2)$ is the proton Dirac form factor. However, since the
nucleon is highly of-mass-shell this assumption might be too
naive~\cite{Gutbrod:1973qr}. Indeed, this prescription underestimates the 
JLAB data for $\sigma_{\rm T}$~\cite{Kaskulov:2008xc} and results in a wrong 
interference pattern between \textsc{l/t} components.

A way to model an intermediate  state which is highly off-mass-shell 
is to increase the Fock space available for the virtual nucleon allowing the 
latter to excite into resonances. Similar to the reggeized-exchange,
these resonances with higher masses and spins may lie on the nucleon Regge trajectory
or correspond to higher mass states with the same angular momentum as the nucleon.  
We replace the Born term in the $s$-channel for $\pi^+$ production by a sum
over all resonance excitations
\be
\label{DSsum_s}
\frac{F_s(Q^2,M_p)}{s-M_p^2+i0^+} 
\to \sum \limits_{i} r(M_i) c(M_i)
\frac{F(Q^2,M_i^2)}{s-M_i^2+i0^+}, 
\ee
where the sum runs from the nucleon-pole contribution, $M_i$ is the $i$th
resonance mass, $r$ and $c$ are the electromagnetic and strong couplings, 
respectively, relative to the lowest lying nucleon state, {\it e.g.} 
$r(M_p) c(M_p)=1$. For $\pi^-$ production we use a similar expansion over 
the $u$-channel contributions 
\be
\label{DSsum_u}
\frac{F_u(Q^2,M_p)}{u-M_p^2+i0^+} 
\to \sum \limits_{i} r(M_i) c(M_i) 
\frac{F(Q^2,M_i^2)}{u-M_i^2+i0^+}.
\ee

In the region of interest for the experiments to be discussed later in 
this paper the invariant mass is $W \agt 2$ GeV and thus in a region where 
the DIS regime starts.
In order to make a connection to our earlier work in which we modeled the
transverse cross section by a partonic
subprocess~\cite{Kaskulov:2008xc,Kaskulov:2009gp} 
we now invoke duality for the exclusive processes treated here. We start
with the Bloom-Gilman duality~\cite{Bloom:1970xb} in the local form  
\begin{equation} 
\label{BG}
F_2^p(x_{\rm B},Q^2) = \sum_{i} (M_i^2-M_p^2+Q^2)  W(Q^2,M_i) \delta(s - M_i^2), 
\end{equation}
where $x_{\rm B}$ stands for a Bjorken scaling variable and the deep inelastic
structure function $F_2^p(x_{\rm B},Q^2)$ is expressed as a sum of resonances.
In Eq.~(\ref{BG}) the hadronic basis is used as a substitute for the quark
basis. When $Q^2$ is large the bulk structure of the resonances becomes less 
and less important and we are justified when taking
the zero-width approximation~\cite{Domokos:1971ds}. 
$W(Q^2,M_i)$ defines the $i$th resonance contribution to the $\gamma^* p$ forward
scattering amplitude; it is essentially the electromagnetic coupling constant
$r(M_i)$ times a resonance form factor $F(Q^2,M_i)$ normalized to unity at 
$Q^2=0$~\cite{Elitzur:1971tg}:
\be
W(Q^2,M_i) = r^2(M_i) [F(Q^2,M_i)]^2, ~~ F(0,M_i)=1.
\ee
A resonance with mass $M_i$ contributes to the structure
function at Bjorken $x_i=Q^2/(M_i^2-M_p^2+Q^2)$. 

To be in line with measurements in the DIS region the resonance form factors
$F(Q^2,M_i^2)$ must fall with $Q^2$ at least as fast as the nucleon dipole form
factor. Futhermore, to be consistent with the scaling behavior of deep inelastic
structure functions the cut-off in the dipole transition form factors
must increase as the mass of the resonance is 
increases~\cite{Elitzur:1971tg,Bloom:1971ye}. Therefore we assume~\cite{Domokos:1971ds}
\be
F(Q^2,M_i^2) = \left(\frac{1}{1 +  \xi \frac{ Q^2}{M_i^2}}\right)^{2},
\ee
where the value of $\xi$ is a common average cut-off parameter. This scenario 
suggests a hardening of the resonance form factors with increasing value of 
$M_i$~\cite{Elitzur:1971tg}.

At high energies the level density of resonances $\rho(M_i^2)$ is large and we
can replace the sum in Eq.~(\ref{BG}) over discrete spectrum of resonances by 
a continuous integral 
\be
\label{sum_to_int}
\sum \limits_{i} \to \int \limits_{M_p^2}^{\infty} dM_i^2 \rho(M_i^2).
\ee
This is clearly a rough approximation in the resonance region itself, but it
makes no difference when we restrict ourselves to the experimental data above
the resonance region. Performing the integration over $M_i$ yields
\be
\label{DualF2resF}
F_2^p(x_{\rm B},Q^2) = (s-M_p^2+Q^2)  r^2(s) [ F(Q^2,s)]^2 \rho(s).
\ee

The structure function $F_2^p$ can be written in the form of a polynomial in
$1-1/\omega'$ where $\omega'=1+W^2/Q^2 $ is a Bloom-Gilman variable. 
As $Q^2/W^2 \to \infty$ or $\omega' \to 1$ the leading term yields the 
Drell-Yan-West behavior 
\be
\label{DYW}
F_2^p(\omega') \propto (\omega'-1)^3, 
\ee
which shows that the power law behavior of the form factor is related to the
suppression of the structure functions in the limit where one quark carries
all of the hadron's momentum. The approximation (\ref{DYW}) is supposed to be reasonable down
to $x_{\rm B} \simeq 0.2$.
Expanding the resonance form factors for $\omega'\to 1$
the leading term reads
\be
F(Q^2,s) = \frac{(\omega'-1)^2}{\xi^2} + \mathcal{O}((\omega'-1)^3). 
\ee
The duality relation, Eq.~(\ref{DualF2resF}), can be written in the form 
\be
\left({\omega'-1}\right)^3
\propto Q^2  \frac{(\omega'-1)^4}{\xi^4} r^2(s) \rho(s). 
\ee
This translates into
\be
r^2(s) \rho(s) \propto \frac{1}{Q^2 (\omega'-1)} = \frac{1}{s}.
\ee
Since the level density grows with increasing $s$, for instance,
$ \rho \propto \exp(const \times M_i) $,
the coupling strength to resonances is decreasing, {\it i.e}, $
r(s_i) \propto (s_i \rho(s_i))^{-1/2} $ where $s_i=M_i^2$.
This simple result has a remarkable consequence. Although an infinite tower of
resonances can contribute to the structure function the weight of resonances
decreases as $1/s$ with increasing value of $s$. 

A vanishing coupling of the higher
spin (mass) resonances to $\pi N$ is expected from the chiral
phenomenology~\cite{Glozman:2007jt}. The latter claim is consistent with our
observation that the more we excite the nucleon the less it decays into the exclusive channel.
Assuming for the strong coupling a similar form 
$
c(s_i) \propto (s_i^{(2\beta-1)} \rho(s_i))^{-1/2}$ with $\beta \ge 1$
the integration in Eqs.~(\ref{DSsum_s}) and~(\ref{DSsum_u}) is superconvergent and
can be carried out analytically. 
Without an explicit assumption about the behavior of the level density 
we get the following form for the product
\be
\label{DS}
\rho(s_i) \, r(s_i) c(s_i) = \frac{1}{\lambda} s_i^{-\beta},
\ee
where $\lambda$ is a normalization constant and $\beta \ge 1$ 
accounts for the behaviour of coupling constants as well as a deviation of the
level density contributing to the exclusive channel compared to the total
inclusive density of states.

We now absorb all the effects of the higher lying resonances into the
nucleon-pole term by setting
\ba
\sum \limits_{i} r(M_i) c(M_i)
\frac{F(Q^2,M_i^2)}{s-M_i^2+i0^+} \Longrightarrow \hspace{2.cm}\nonumber \\
\int \limits_{M_p^2}^{\infty} dM_i^2 \rho(M_i^2) r(M_i^2)c(M_i^2)
\frac{F(Q^2,M_i^2)}{s-M_i^2+i0^+} \nonumber \\
= \int \limits_{M_p^2}^{\infty} ds_i \frac{s_i^{-\beta}}{\lambda}
\frac{F(Q^2,s_i)}{s-s_i+i0^+} \equiv \frac{F_s(Q^2,s)}{s-M_p^ 2+i0^+},
\ea
where the sum in Eq.~(\ref{DSsum_s}) over discrete spectrum of resonances 
has been replaced again by a continuous integral. $F_s(Q^2,s)$ is the form
factor on the {\it r.h.s.} of Eq.~(\ref{F1generic}). Similarly we proceed for 
the transition form factor $F_u(Q^2,u)$ in the $u$-channel, Eq.~(\ref{DSsum_u}). 
The integration covers the full region from the nucleon pole $M_p$ up to
$\infty$.   
Furthermore, the normalization constants $\lambda$ are determined by the
charge conservation at the real photon point $Q^2=0$, {\it i.e.} 
\ba
\label{Lambda_s_ch}
\lambda\Big|_{s-channel} = (s-M_p^2) \int \limits_{M_p^2}^{\infty} ds_i \,  
\frac{\rho(s_i)\, r(s_i)c(s_i)}{s-s_i+i0^+}, \\
\label{Lambda_u_ch}
\lambda\Big|_{u-channel} = (u-M_p^2) \int \limits_{M_p^2}^{\infty} ds_i \,  
\frac{\rho(s_i)\, r(s_i)c(s_i)}{u-s_i^2+i0^+},
\ea
for the $s$- and $u$-channels, respectively. This merely  guarantees
that the effective form factors are normalized to unity,
{\it e.g.} $F_{s(u)}(0,s(u))=1$. With this prescription we demand that the contributions of 
resonances show up in the modified off-mass-shell behavior of the nucleon
transition form factors.

The $s$- and $u$-channel form factors read 
\begin{eqnarray}
\label{F1BGres_s_ch_beta4} 
{F}_{s}(Q^2,s) &=&  
\frac{ \ds
\int \limits_{M^2_p}^{\infty} ds_i 
\frac{s_i^{-\beta}}{s-s_i+i0^+}  \left(\frac{1}{1 +  \xi \frac{ Q^2}{s_i}}\right)^{2}
}
{\ds \int \limits_{M_p^2}^{\infty} ds_i   \,  \frac{s_i^{-\beta}}{s-s_i+i0^+} },
\\
\label{F1BGres_u_ch_beta4}
{F}_{u}(Q^2,u) &=& 
\frac{ \ds
\int \limits_{M^2_p}^{\infty} ds_i 
\frac{s_i^{- \beta} }{u-s_i+i0^+} \left(\frac{1}{1 +  \xi \frac{ Q^2}{s_i}}\right)^{2}
}
{\ds \int \limits_{M_p^2}^{\infty} ds_i   \,  \frac{s_i^{-\beta}}{u-s_i+i0^+} },
\end{eqnarray}
Because of the singularity at $s_i=s+i0^+$ the
$s$-channel integrals develop an imaginary part which is missing
in the $u$-channel contribution where the branch point sits in the unphysical region.

\begin{figure}[b]
\begin{center}
\includegraphics[clip=true,width=1\columnwidth,angle=0.]
{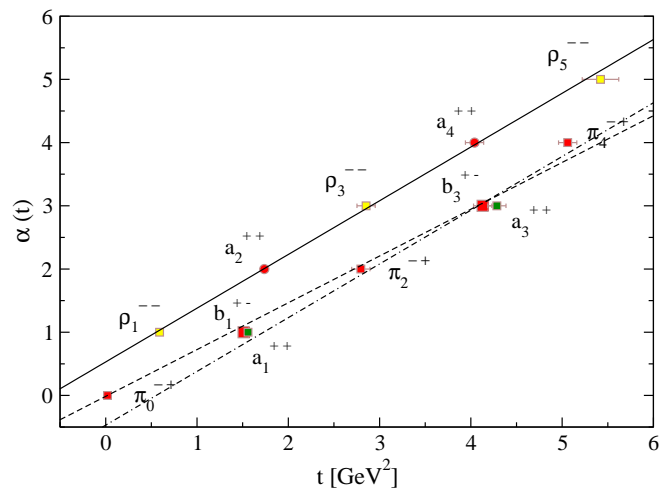}
\caption{\label{ReggeTr} 
\small (Color online) $\rho(770)/a_2(1320)$ (solid), $\pi/b_1(1235)$ (dashed) and $a_1(1260)$
(dash-dotted) Regge trajectories.
\vspace{-0.4cm}
}
\end{center}
\end{figure}

Concerning the terminology for regions like the one at JLAB it would be appropriate
to use the words {\it resonance} effect. On the other hand, in the DIS region at
HERMES it would be more natural to describe the effect as of
{\it partonic} origin. Since both descriptions are dual in the context of the present
approach we shall in line of~\cite{Kaskulov:2008xc,Kaskulov:2009gp} 
refer to the terms derived above as the {\it resonance/partonic} (\textsc{r/p}) contributions.

\section{Vector and axial-vector Regge trajectories}
The mesonic Regge trajectories can be characterized by the signature and parity.
The signature determines whether the Regge poles in the scattering amplitude
will occur for even or odd positive integer value of the 
trajectory $\alpha(t)=J$~(the spin). The leading mesons contributing to $(e,e'\pi^{\pm})$ 
are the natural $P=(-1)^{J}$ parity vector $\rho(770)$ and the unnatural
$P=(-1)^{J+1}$ parity axial-vector mesons $a_1(1260)$ and $b_1(1235)$. The
Regge  trajectories $\alpha(t)$ considered here are shown in Figure~\ref{ReggeTr}

The absolute contribution of the reggeized $\rho$-exchange amplitude to 
$(e,e'\pi^{\pm})$  turns out to be small, but by its interference with the 
$s$- and $u$-channel terms considered above, it is responsible
for the $\pi^-/\pi^+$ asymmetry in photoproduction and gives a sizable
contribution to the $\pi^-/\pi^+$ ratio in electroproduction.

In the axial-vector sector, the experimental isolation of the amplitudes with
axial-vector quantum numbers would be
of great interest. For instance, using a proton target polarized perpendicular
to the reaction plane, and photons polarized parallel to the reaction plane,
one can directly access the difference between recoil and the polarized target
asymmetries which is proportional to the exchange of the 
$a_1(1260)$-trajectory~\cite{Worden:1972dc}. However, being suppressed by the Regge factor
$\sim e^{-\alpha_{a_1}'\ln(\alpha_{a_1}'s)m^2_{a_1}}$ at $t=0$, its contribution to
the forward unpolarized cross section turns out to be small.
With our choice of the $b_1NN$ tensor coupling the contribution of $b_1(1235)$ exchange is
even smaller.  On the other hand, as we shall see, it is absolutely essential
to consider the exchange of $a_1(1260)$ Regge trajectory in the polarization $(\vec{e},e'\pi)$
observables like the beam spin azimuthal asymmetry considered in the
following. Other aspects related to a possible role of $a_1(1260)$ in $(e,e'\pi)$ are
discussed in \cite{Bechler:2009me}.

\begin{widetext}

\subsection{Vector-isovector $I^G(J^{PC})=1^+(1^{--})$ exchange currents}

The currents $J_{\rho}^{\mu}$ describing the exchange of the natural parity 
$\rho(770)$-meson Regge trajectory are given by
\begin{eqnarray}
\label{rho}
\left[
\begin{array}{c}
-iJ_{\rho}^{\mu}(\gamma^*p\to\pi^+n) \\ \\
-iJ_{\rho}^{\mu}(\gamma^*n\to\pi^-p)
\end{array}
\right]
= - i
\
\sqrt{2} G_{\rho\gamma\pi}
\, G_{\rho NN}
F_{\rho\gamma\pi}(Q^2)
{\varepsilon^{\mu\nu\alpha\beta} q_{\nu}
 k_{\alpha}}  
\bar{u}_{s'}(p')
\left[  (1 + \kappa_{\rho}) \gamma_{\beta}
- \frac{\kappa_{\rho}}{2M_p} (p+p')_{\beta} \right] u_s(p) \nonumber \\
\times \left[\frac{1 - e^{-i\pi\alpha_{\rho}(t)}}{2}\right]
\left(- \alpha'_{\rho} \right) \Gamma[1-\alpha_{\rho}(t)]
e^{\ln(\alpha'_{\rho}s)(\alpha_{\rho}(t)-1)}
\end{eqnarray}
where $G_{\rho NN}=3.4$ and $\kappa_{\rho}=6.1$ are the standard vector and 
anomalous tensor coupling constants, respectively. The $\rho$-trajectory 
adopted here reads $\alpha_{\rho}(t) = 0.53+\alpha_{\rho}'t$  
with a slope $\alpha_{\rho}'=0.85$~GeV$^{-2}$. The $\Gamma$ function in
(\ref{rho}) contains the pole propagator $\sim 1/\sin(\pi\alpha_{\rho}(t))$ 
but no zeroes and the amplitude zeroes only occur through the factor 
$1 - e^{-i\pi\alpha_{\rho}(t)}$.

The ${\rho\gamma\pi}$ coupling constants $G_{\rho\gamma\pi}$ can be deduced
from the radiative $\gamma\pi$ decay widths of $\rho$
\be
\label{Vgammapi}
\Gamma(\rho^{\pm} \to \gamma \pi^{\pm}) = \frac{\alpha_e}{24}
\frac{G_{\rho \gamma\pi}^2}{m_{\rho}^3} 
\left({m_{\rho}^2} - {m_{\pi}^2}\right)^3.
\ee
The measured width~\cite{PDG}:
$
\Gamma_{\rho^{\pm}\to \gamma\pi^{\pm}} = (68 \pm 7)\mbox{~keV},
$
where the central value corresponds to $G_{\rho\gamma\pi} = 0.728$ GeV$^{-1}$.
For the transition form factor $F_{\rho\gamma\pi}(Q^2)$ we use a VMD model 
$F_{\rho\gamma\pi}(Q^2)=(1+Q^2/\Lambda^2_{\rho\gamma\pi})^{-1}$ with 
$\Lambda_{\rho\gamma\pi}=m_{\omega(782)}$.

\subsection{Axial-vector $I^G(J^{PC})=1^-(1^{++})$ exchange currents}
The axial-vector $a_1(1260)$ meson with  $I^G(J^{PC})=1^-(1^{++})$ 
has a large width into the $a_1(1260)^{\pm}\to\gamma\pi^{\pm}$ channel~\cite{PDG}.
A conversion of $a_1$ into $\gamma\pi$ is described by the $a_1\gamma\pi$ 
vertex  $\mathcal{L}_{a_1\gamma\pi}=\frac{-i e}{4} G_{a_1\gamma\pi} F^{\mu\nu} 
\langle Q [A_{\mu\nu},\varphi] \rangle$ where $F^{\mu\nu}$ denotes the field
tensor of photons, $A_{\mu\nu}$ stands
for the field tensor of the axial-vector meson with
$A_{\mu\nu}= \partial_{\mu}A_{\nu}-\partial_{\nu}A_{\mu}$ and
\be
A_{\mu}=
\left(
\begin{array}{ccc}
a_1^0  &     \sqrt{2}a_1^+ \\
    \sqrt{2}a_1^-       & - a_1^0
\end{array}
\right)_{\mu}.
\ee
$\varphi$ is a
standard $SU(2)$ pion matrix, $Q=\mbox{diag}(2/3,-1/3)$ is a quark charge
matrix, $\langle .. \rangle $ and $[..]$ denote a trace and a commutator of
fields. 
The hadronic $a_1NN$ interaction is described by
\be
\mathcal{L}_{a_1NN} = G_{a_1NN} \bar{\psi}   \gamma^{\mu} \gamma_{5}
 A_{\mu} \psi,
\ee
where $\psi=(p,n)^{\rm T}$. Because of $G$-parity conservation 
in the vertex there is no tensor coupling of $a_1$ to nucleons. 

In the reactions $p(\gamma^*,\pi^{+})n$ and $n(\gamma^*,\pi^{-})p$
the currents describing the exchange of $a_1(1260)$ trajectory read
\begin{eqnarray}
\left[
\begin{array}{l}
-iJ_{a_1}^{\mu}(\gamma^*p\to\pi^+n) \\ \\
-iJ_{a_1}^{\mu}(\gamma^*n\to\pi^-p)
\end{array}
\right]
= \left[
\begin{array}{l}
+ \\ \\
-
\end{array}
\right]
\sqrt{2}
 \, G_{a_1 NN} G_{a_1\gamma\pi} F_{a_1\gamma\pi}(Q)
\Big[  k^{\mu} q^{\nu} -(qk)g^{\mu\nu}\Big] 
\bar{u}_{s'}(p') \gamma_{\nu} \gamma_{5} u_s(p)  \nonumber \\
\times \left[\frac{1 - e^{-i\pi\alpha_{a_1}(t)}}{2}\right]
\left(- \alpha'_{a_1} \right) \Gamma[1-\alpha_{a_1}(t)]
e^{\ln(\alpha'_{a_1}s)(\alpha_{a_1}(t)-1)},
\end{eqnarray}
The $a_1$ Regge trajectory adopted here is
$\alpha_{a_1}(t) = \alpha_{\rho}(t)-1$  
where $\alpha_{\rho}(t)$ is the trajectory of $\rho$.
The $\gamma^{*}\to a_1 \pi$ transition is isovector and contributes
with opposite signs to $\gamma^*p\to\pi^+n$ and $\gamma^*n\to\pi^-p$ reactions.

To estimate the $a_1$-nucleon coupling constant $G_{a_1NN}$ one can relate say $G_{a_1 pp}$
to the observed axial-vector coupling constant using axial-vector
dominance~\cite{SweigWada,Birkel:1995ct}
$\frac{g_A}{g_V} = \frac{\sqrt{2} f_{a_1} G_{a_1 pp}}{m_{a_1}^2}$,
where the weak decay constant $f_{a_1}$ is deduced from $\tau$ decay:
$\tau \to a_1 +\nu_{\tau}$.
With ${g_A}/{g_V}=1.267, $ $f_{a_1} = (0.19\pm0.03)$~GeV$^2$ one gets the
following estimate 
$G_{a_1 pp}=G_{a_1 NN}=7.1\pm 1.0$.

The radiative decay width $a_1\to \gamma\pi$ is given by
\be
\label{GammaA1}
\Gamma_{a_1^+\to \gamma\pi^+} =  \frac{\alpha_e}{24}
\frac{G_{a_1\gamma\pi}^2}{ m_{a_1}^3} (m_{a_1}^2-m_{\pi}^2)^3,
\ee
The empirical width $a_1^{+}\to \gamma\pi^{+}$ is
$\Gamma_{a_1^+\to \gamma\pi^+ } \simeq (640\pm 246)~\mbox{keV}$~\cite{PDG}.
The coupling constant $G_{a_1\gamma\pi} \simeq 1.1$~GeV$^{-1}$ corresponds
to the central value.
In a VMD picture a conversion of $\gamma$ to $\rho$
with subsequent $a_1\rho\pi$ interaction generates the
monopole form factor
$F_{a_1\gamma\pi}(Q^2) = (1+Q^2/\Lambda^2_{a_1\gamma\pi})^{-1}$ with $\Lambda_{a_1\gamma\pi}=m_{\rho(770)}$.
This form is used to model the $Q^2$ dependence of the $a_1\gamma^*\pi$
vertex.

\subsection{Axial-vector $I^G(J^{PC})=1^+(1^{+-})$ exchange currents}
We consider the exchange of $b_1(1235)$ axial-vector meson with  $I^G(J^{PC})=1^+(1^{+-})$.
The conversion of $b_1 \to \gamma\pi$ is described by the vertex
$
\mathcal{L}_{b_1\gamma\pi} = \frac{e G_{b_1\gamma\pi}}{4} F^{\mu\nu} 
\langle Q \{B_{\mu\nu},\varphi\} \rangle,
$
where $\{..\}$ anti-commutes and $B_{\mu\nu}=\partial_{\mu}
B_{\nu}-\partial_{\nu} 
B_{\mu}$ with
\be
B_{\mu}=
\left(
\begin{array}{ccc}
b_1^0 &     \sqrt{2}b_1^+ \\
    \sqrt{2}b_1^-       & - b_1^0
\end{array}
\right)_{\mu}.
\ee
The $b_1(1235)$ coupling to nucleons takes the form of axial-tensor interaction
\begin{eqnarray}
\label{bNN}
\mathcal{L}_{b_1 NN} &=& i \frac{G_{b_1 NN}}{4M_N} \,\bar{\psi}
\sigma^{\mu\nu}\gamma_5 B_{\mu\nu}\psi.
\end{eqnarray}
where $\sigma^{\mu\nu}=\frac{i}{2}[\gamma^{\mu},\gamma^{\nu}]$. 
The hadronic currents $-iJ_{b_1}^{\mu}$ describing the exchange of $b_1(1235)$ Regge
trajectory read
\begin{eqnarray}
\left[
\begin{array}{l}
-iJ_{b_1}^{\mu}(\gamma^*p\to\pi^+n) \\ \\
-iJ_{b_1}^{\mu}(\gamma^*n\to\pi^-p)
\end{array}
\right] 
&=& 
\frac{\sqrt{2}}{3}
 \, \frac{G_{b_1 NN}}{2M_N} G_{b_1\gamma\pi} F_{b_1\gamma\pi}(Q)
\Big[  k^{\mu} q^{\nu} -(qk)g^{\mu\nu}\Big] (p+p')_{\nu} 
\bar{u}_{s'}(p') \gamma_{5} u_s(p) \nonumber \\
& \times & \left[\frac{1 - e^{-i\pi\alpha_{b_1}(t)}}{2}\right]
\left(- \alpha'_{b_1} \right) \Gamma[1-\alpha_{b_1}(t)]
e^{\ln(\alpha'_{b_1}s)(\alpha_{b_1}(t)-1)}.
\end{eqnarray}
The radiative 
decay width of $b_1^{\pm}\to \gamma \pi^{\pm}$ is
$\Gamma_{b_1^{\pm}\to \gamma \pi^{\pm}} = (230\pm60)~\mbox{keV}$~\cite{PDG}.
Making use of an expression similar to Eq.~(\ref{GammaA1}) one gets
$G_{b_1\gamma\pi}/3=0.647$~GeV$^{-1}$. 
The $\pi$ and $b_1(1235)$ Regge trajectories 
 are nearly degenerate (dashed curve if Figure~\ref{ReggeTr}). 
Therefore we assume $\alpha_{b_1}(t)=\alpha_{\pi}(t)$.
In the $b_1\gamma^*\pi$ vertex we use the VMD form factor 
$F_{b_1\gamma\pi}(Q^2) = (1+Q^2/\Lambda_{b_1\gamma\pi}^2)^{-1}$ with $\Lambda_{b_1\gamma\pi}=m_{\omega(782)}$.
\end{widetext}

\begin{figure*}[t]
\begin{center}
\includegraphics[clip=true,width=2\columnwidth,angle=0.]{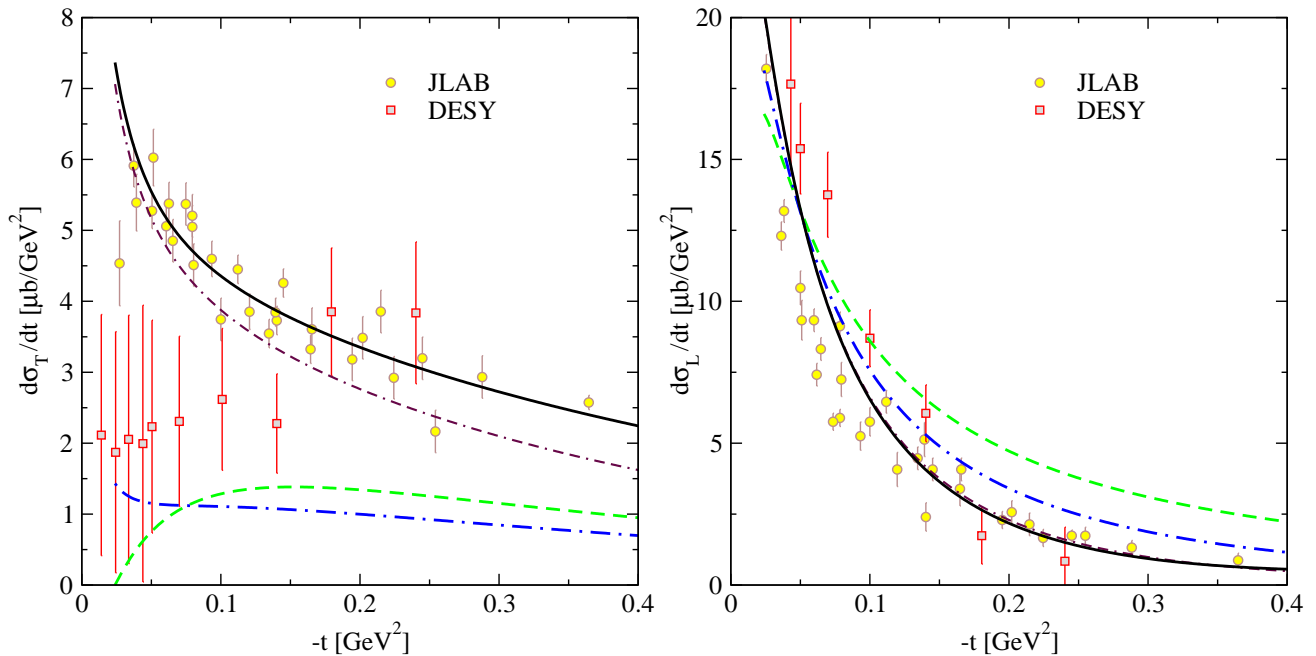}
\caption{\label{PiPdsdtGlobal} 
\small (Color online) $-t$ dependence of the transverse $d\sigma_{\rm T}/dt$
(left panel) and longitudinal $d\sigma_{\rm L}/dt$ (right panel) differential cross sections
in exclusive reaction $p(\gamma^*,\pi^+)n$.
The compilation of JLAB and old DESY data are from  Ref.~\cite{Blok:2008jy} and have been
scaled to common values of $W=2.19$~GeV and $Q^2=0.7$~GeV$^2$. 
The solid curves describe the model results which include the
effect of resonances and exchange of $\pi/b_1(1235)$, $\rho(770)/a_2(1320)$ and  $a_1(1260)$ 
Regge trajectories. The same is true for the dash-dash-dotted curves but without the
DIS slope of Eq.~(\ref{DISslope}).
The dashed curves correspond to the contribution of the $\pi$-reggeon exchange only. 
The dash-dotted curves describe the model results which include the exchange of
Regge trajectories and on-mass-shell parameterization of the proton Dirac form
factor. 
\vspace{-0.4cm}
}
\end{center}
\end{figure*}

It was proposed~\cite{Guidal:1997hy} that the polarized photon asymmetry in
$p(\gamma,\pi^0)p$ reaction at high energies
can be used to estimate the product of the $b_1$ electromagnetic and strong coupling constants.
In~\cite{Guidal:1997hy} the $b_1^0(1235)$ is coupled to the axial-vector current and
the axial-tensor interaction has been neglected. 
However, the axial-vector $\gamma^{\mu}\gamma_5$ vertex is $C$-parity even and can not
couple to the $b_1^0(1235)$-meson which has negative $C$-parity. This is opposed to the
$a_1(1260)$ which couples to $\gamma^{\mu}\gamma_5$.
By $C(G)$-parity only a tensor $b_1NN$ coupling, Eq.~(\ref{bNN}), is
possible. We have checked and found that with the proper $b_1NN$ vertex this
extraction is not obvious. With the tensor interaction the exchange of the $b_1(1235)$
trajectory is negligibly small and can be
readily neglected. For instance, using
$G_{b_1 NN} = G_{a_1 NN}$ or even increasing considerably the latter value
does not produce any noticeable effects on the observables considered
here.

\section{Electroproduction above the resonance region}

In this section we demonstrate the resonance interpretation proposed in
this work and fix the model parameters using the JLAB data.
At first, we briefly consider the real photon limit of the Regge amplitudes at
high energies. In $\pi^+$ and $\pi^-$ photoproduction at very forward angles the 
reggeized electric amplitudes, see Eqs.~(\ref{PipolePl}) and (\ref{PipoleMi}), 
are supposed to be dominant. Since the vector and axial-vector 
meson-exchange contributions vanish at forward angles, Eqs.~(\ref{PipolePl}) 
and (\ref{PipoleMi}) are parameter free, provided the intercept of the 
$\pi$-trajectory and the $g_{\pi NN}$ coupling constant are fixed. However, 
further assumptions concerning a choice of the phases in the Regge amplitudes 
have to be made. In the gauged $\pi$-Regge amplitudes, Eqs.~(\ref{PipolePl}) 
and~(\ref{PipoleMi}), an assumption concerning an exact degeneracy of
$\pi$ and axial-vector $b_1(1235)$ Regge trajectories with a choice of 
rotating phase in $\pi^+$ and a constant phase in $\pi^-$ photoproduction 
yields a remarkable consistency with data~\cite{Guidal:1997hy}. The degeneracy 
of $\rho(770)/a_2(1320)$ Regge trajectories and $G$-parity arguments result 
in a rotating  phase in $\pi^+$ and a constant phase in $\pi^-$ production described by 
the $\rho$-exchange amplitude, Eq.~(\ref{rho}). Here, to be consistent with the real
photon limit, we follow these 
assumptions~\cite{Guidal:1997hy}. However, in the high-$Q^2$ electroproduction 
a particular choice of phases in the Regge amplitudes is of minor importance. The 
virtual photon $(\gamma^*,\pi^{\pm})$ results presented here can be well reproduced with 
the standard Regge propagators. From the meson spectrum there is no conclusive
evidence that a leading Regge trajectory for an unnatural parity $\rho_2$ state
exists. Therefore, we do not make any assumptions on a  
degeneracy pattern of $a_1(1260)$.

The resulting Regge model based on reggeized gauge invariant Feynman  
amplitudes describes  the high energy  $\pi^{\pm}$ photoproduction data 
relevant for the present studies  reasonably well. These include the 
differential  cross  sections  above  the   resonance region, the   
$\pi^-/\pi^+$ ratio of the $n(\gamma,\pi^-)p$ and $p(\gamma,\pi^+)n$  
differential cross sections and polarized photon asymmetries. Our 
description of photoproduction data  is  quantitatively similar to the 
results of Ref.~\cite{Guidal:1997hy} and we do not repeat this comparison 
with experimental data here.  

An extension of the model to electroproduction is straightforward provided 
the $Q^2$-dependent transition form factors are defined. In 
Figure~\ref{PiPdsdtGlobal} we plot the transverse $d\sigma_{\rm T}/dt$ 
(left panel) and longitudinal $d\sigma_{\rm L}/dt$ (right panel)  $\pi^+$ 
electroproduction data from JLAB and DESY  (old data) scaled to the same 
values of $Q^2=0.7$~GeV$^2$ and $W=2.19$~GeV~\cite{Blok:2008jy}. The 
value of the momentum cut-off $\Lambda_{\gamma\pi\pi}$  in the  pion form 
factor, Eq.~(\ref{PiFF}),  is largely  constrained  by  the magnitude of 
$d\sigma_{\rm L}/dt$ at forward angles. In the following the JLAB data are 
considered to be a guideline for fixing the model parameters. The dashed 
curves correspond to $\Lambda_{\gamma\pi\pi}^2=0.46$~GeV$^2$ and describe the 
contribution of the $\pi$-reggeon exchange only. The exchange of $\pi$ 
dominates in $d\sigma_{\rm L}/dt$ at forward angles. However, in 
$d\sigma_{\rm T}/dt$  the $\pi$-exchange is not compatible with data and 
also vanishes in the forward direction. The dash-dotted curves correspond 
to the gauged electric model with the on-shell Dirac form factor, see 
Eq.~(\ref{FsF1p}),
\be
\label{F1p}
{F}_1^p(Q^2) = 
\frac{\ds {G_{E}^p(Q^2)}/{G_{M}^p(Q^2)} + {Q^2}/{4M_p^2}}
{\ds 1+ {Q^2}/{4M_p^2}}G_M^p(Q^2),
\ee
and exchange of $\rho(770)/a_2(1320)$ and $a_1(1260)$ Regge trajectories. In Eq.~(\ref{F1p}) 
the electric form factor $G_{E}^p$ decreases linearly as a function of $Q^2$ 
with respect to the magnetic form factor $G_{M}^p$ with a node around
$Q^2_0\simeq 8$~GeV$^2$~\cite{Kaskulov:2003wh} provided 
${\mu_p G_E^p}/{G_M^p}= 1-{Q^2}/{Q_0^2}$. On the other hand up to 
$\simeq 5$~GeV$^2$ the magnetic form factor is a dipole 
$G_{M}^p = \mu_p/(1+{Q^2}/0.71 \mbox{GeV}^2)^{2}$ where $\mu_p=2.793$ is the 
magnetic moment of the proton. 

\begin{figure}[t]
\begin{center}
\includegraphics[clip=true,width=1\columnwidth,angle=0.]
{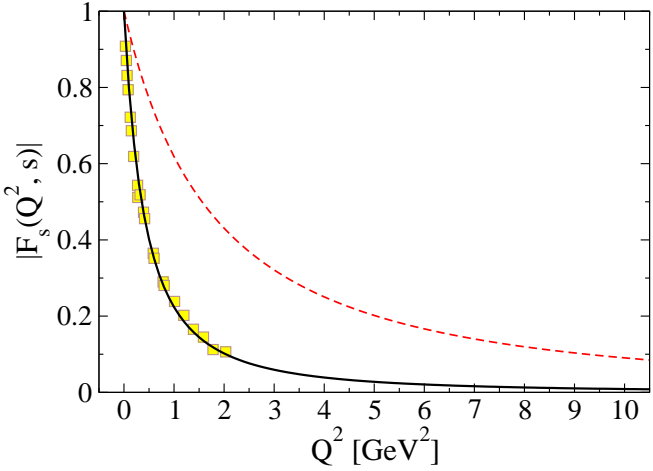}
\caption{
\label{F1onoff} 
(Color online) The $Q^2$ dependence of the absolute value 
of the transition form factor $|F_s(Q^2,s)|$ (dashed curve), 
Eq.~(\ref{F1_s_res}), at $\sqrt{s}=2.2$~GeV. The solid curve 
describes the proton Dirac form factor,  Eq~(\ref{F1p}), in 
comparison with data. \vspace{-0.4cm}
}
\end{center}
\end{figure}

As one can see, this model (dash-dotted curves) with the nucleon-pole 
(gauge invariance), vector and axial-vector meson-exchange Regge trajectories 
describes  $d\sigma_{\rm L}/dt$ well and grossly underestimates $d\sigma_{\rm
  T}/dt$. Variations  of the cut-offs in the vector and axial-vector meson 
transition form factors do not improve the description. This preliminary 
comparison  with data shows that being consistent with  photoproduction  
the above simple extension of the Regge model to electroproduction is not 
able to describe the data already at values of $Q^2$ as low as $Q^2 \simeq
1$~GeV$^2$. The discrepancies with data increase with increasing value of 
$Q^2$~\cite{Blok:2008jy}.

\begin{table*}
\begin{tabular}{c|c|c|c|c}
\hline
\hline
Regge  & Parameters & $p(\gamma^*,\pi^+)n$ & 
$n(\gamma^*,\pi^-)p$ & Regge trajectory  \\ 
exchange     &   &   &   & 
$\alpha_{i}(t)=\alpha^0_i+\alpha'_i\,t$ \\
\hline
$\pi(140)/b_1(1235)$  & $g_{\pi NN}=13.4$ & 
$+e^{-i\pi\alpha_{\pi}(t)}$ & $1$&
$\alpha_{\pi}(t) = \alpha'_{\pi} (t-m_{\pi}^2)$\\
       & 
   &
  & & $\alpha'_{\pi}=0.74$ \\
\hline 
 & $G_{\rho NN}=3.4$  &  & &  \\
$\rho(770)/a_2(1320)$ &  $\kappa_{\rho}=6.1$ & 
$- e^{-i\pi\alpha_{\rho}(t)}$ & 
$1$ & $\alpha_{\rho}(t) = 0.53+0.85t$ \\
       &  $G_{\rho\gamma\pi} = 0.728~\mbox{GeV}^{-1}$ &   &  &   \\
 & $\Lambda_{\rho\gamma\pi}=m_{\omega(782)}$  &  &  \\
\hline 
 & $G_{a_1 NN}=7.1$  &  & & \\
$a_1(1260)$ & $G_{a_1\gamma\pi} = 1.1~\mbox{GeV}^{-1}$  & 
$\frac{1 - e^{-i\pi\alpha_{a_1}(t)}}{2}$ & $\frac{1 -
  e^{-i\pi\alpha_{a_1}(t)}}{2}$ 
 & $\alpha_{a_1}(t)=\alpha_{\rho}(t)-1$ \\
&  $\Lambda_{a_1\gamma\pi}=m_{\rho(770)}$  &  &  \\
\hline 
\hline 
 &   &   &  & $\alpha_{\pi}(t)$ \\
Resonances & $\beta=3$, $\xi=0.4$  & $+e^{-i\pi\alpha_{\pi}(t)}$ & 
1 & $\mbox{with}~\alpha'_{\pi} \to
\frac{\alpha_{\pi}'}{1+a\frac{Q^2}{W^2}}$ \\
&    &  &  & $a=2.4$ \\
\hline
\hline
\end{tabular}
\caption{
\label{TableMP}
A summary table of a model parameters. See the text for the details. 
\vspace{-0.0cm}
}
\end{table*}

Next, consider the resonance contributions using the transition 
form factors as defined in Eq.~(\ref{F1BGres_s_ch_beta4}). 
We have two parameters at hand: the parameter $\beta$ which is related 
to the level density of states and a parameter $\xi$ describing the average 
cut-off in the resonance transition form factors. All the exclusive 
$p(\gamma^*,\pi^+)n$ and $n(\gamma^*,\pi^-)p$ electroproduction data 
considered in this work from JLAB, DESY and Cornell to DIS region at HERMES 
can be well described by the choice $\beta=3$ and $\xi=0.4$. 
The formulae for the transition form factors given in
(\ref{F1BGres_s_ch_beta4}) and (\ref{F1BGres_u_ch_beta4}) can be integrated
and yield (for $\beta=3$)
\begin{widetext}
\be
\label{F1_s_res}
{F}_{s}(Q^2,s) = 
\frac{\ds
s \ln\left[\frac{ \xi Q^2}{M_p^2}+1\right] 
\frac{ (2\xi Q^2+s)}{(\xi Q^2)^2} 
- \frac{s (\xi Q^2+s) }{ \xi Q^2 \,(\xi Q^2+M_p^2)} 
+ \ln\left[
  \frac{s-M_p^2}{M_p^2}\right] -i\pi}
{\ds \left(\frac{\xi Q^2}{s}+1\right)^2 \left( \frac{
    s^2+2s M_p^2}{ 2 M_p^4} + \ln\left[\frac{ s-M_p^2}{ M_p^2}\right]-i\pi \right)},
\ee
\be
\label{F1_u_res}
{F}_{u}(Q^2,u) = 
\frac{\ds
u \ln\left[\frac{ \xi Q^2}{M_p^2}+1\right] 
\frac{ (2\xi Q^2+u)}{(\xi Q^2)^2} 
- \frac{u (\xi Q^2+u) }{ \xi Q^2 \, (\xi Q^2+M_p^2)} 
+ \ln\left[
  \frac{M_p^2-u}{M_p^2}\right]}
{\ds \left(\frac{\xi Q^2}{u}+1\right)^2 \left( \frac{
    u^2+2u M_p^2}{ 2 M_p^4} + \ln\left[\frac{M_p^2-u}{ M_p^2}\right]\right)}.
\ee

\end{widetext}
In Figure~\ref{F1onoff} we plot 
the $Q^2$ dependence of the absolute value 
of the transition form factor $|F_s(Q^2,s)|$ (dashed
curve) at $W=\sqrt{s}=2.2$~GeV in comparison with the on-shell
parameterization of the proton's Dirac form factor $F_1^ p(Q^ 2)$
Eq.~(\ref{F1p}) (solid curve). It is clearly seen that $F_s$ is 
considerably harder than $F_1^p$. This 
difference reflects the influence of the higher lying resonances.

There is an additional effect which we would like to take into account.
In Refs.~\cite{Kaskulov:2008xc,Kaskulov:2009gp} it has been observed that 
in exclusive reaction $(e,e'\pi^{+})$ the slope of partonic contributions
 which is driven by the intrinsic transverse momentum distribution of partons 
slightly decreases with increasing value of $Q^2$. Since in our present 
description the contribution of resonances is dual to direct partonic 
interaction we accommodate this anti-shrinkage effect in the transition 
form factors $\mathcal{F}_{s(u)}$,  see Eq.~(\ref{F1generic}), using the 
slope parameter 
\be
\label{DISslope}
\alpha'_{\pi} \to \frac{\alpha'_{\pi}}{1+a \frac{Q^2}{W^2}},
\ee
with $a\simeq 2.4$. This behavior has been found from the fit to the $(Q^2,W)$ 
dependence of the transverse partonic DIS slope of~\cite{Kaskulov:2008xc,Kaskulov:2009gp}. 
Eq.~(\ref{DISslope}) is effective in electroproduction and in the resonance
transition form factors $\mathcal{F}_{s(u)}$, Eq.~(\ref{F1generic}), only. For
real photons the phase of $\mathcal{F}_{s(u)}$ is that of
$\mathcal{F}_{\gamma\pi\pi}$ and a proper Regge limit of~\cite{Guidal:1997hy} is guaranteed.

Then using the \rp transition form factor, 
Eq.~(\ref{F1_s_res}), the transverse cross section 
$d\sigma_{\rm T}/dt$ gets large (solid curve) in agreement with 
JLAB data, see Figure~\ref{PiPdsdtGlobal}. The effect of resonances 
is much smaller in the longitudinal response $d\sigma_{\rm L}/dt$ 
but it improves the description of data at higher values of $-t$. 
As we shall see, the same effect will strongly influence the interference 
cross sections and allow to explain both the sign and magnitude of 
$d\sigma_{\rm TT}/dt$ and $d\sigma_{\rm LT}/dt$. The solid curves include the 
effect of the DIS slope, Eq.~(\ref{DISslope}). The dash-dash-dotted curves 
correspond to the results without Eq.~(\ref{DISslope}); the effect is 
rather small at forward angles and could be partially absorbed in a
redefinition of $\xi$. To be in line with~\cite{Kaskulov:2008xc,Kaskulov:2009gp} 
we keep this phenomenological behavior.

The model parameters are summarized in Table~\ref{TableMP}. The Regge phase pattern
discussed above and used in the calculations is also shown for different
reggeon exchange contributions. The cut-off $\Lambda_{\gamma\pi\pi}$ in the
pion form factor, Eq.~(\ref{PiFF}),
  is a fit parameter. From the fit to the longitudinal data we observe
  essentially three
  regions. At small values of $Q^2<0.4$~GeV$^2$ the model results are remarkably
  consistent with a VMD value of
  $\Lambda^2_{\gamma\pi\pi}=m_{\rho(770)}^2\simeq 0.59$~GeV$^2$. The intermediate region
  $0.6<Q^2<1.5$~GeV$^2$ in $F\pi$-1 experiment~\cite{Tadevosyan:2007yd} 
 demands somewhat smaller value of 
 $\Lambda^2_{\gamma\pi\pi} \simeq 0.4$~GeV$^2$. In the deep $(Q^2,W)$ region
 the JLAB, Cornell and DESY data can be well described using 
$\Lambda^2_{\gamma\pi\pi} \simeq 0.46$~GeV$^2$. In our calculations we shall follow these
  prescriptions for $\Lambda_{\gamma\pi\pi}$.

\begin{figure*}
\begin{center}
\includegraphics[clip=true,width=2\columnwidth,angle=0.]{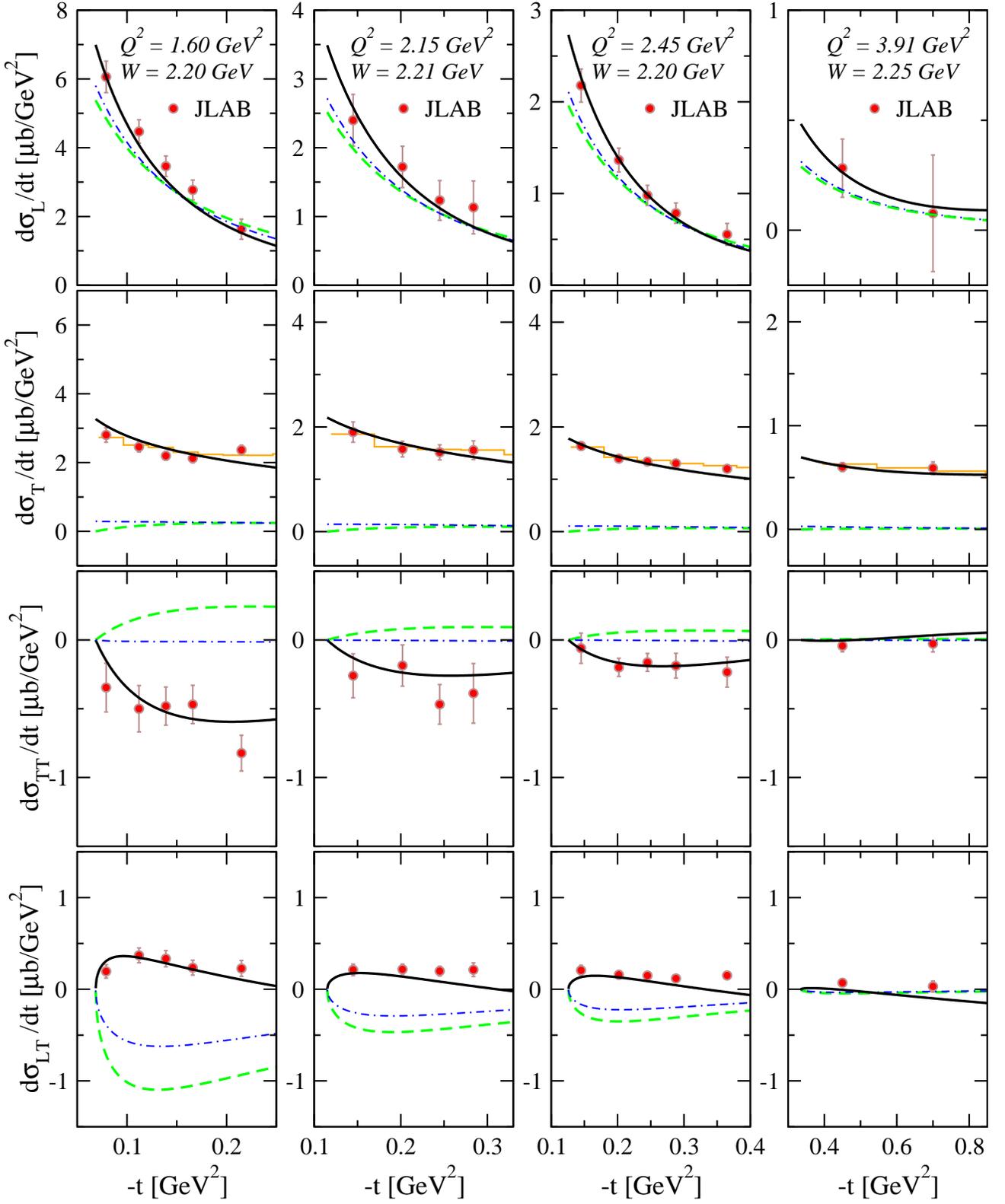}
\caption{\label{Horn1} 
\small (Color online) $-t$ dependence of \textsc{l/t} partial transverse 
$d\sigma_{\rm T}/dt$, longitudinal $d\sigma_{\rm L}/dt$ and interference 
$d\sigma_{\rm TT}/dt$ and $d\sigma_{\rm LT}/dt$ differential cross sections 
in exclusive reaction $p(\gamma^*,\pi^+)n$. The experimental data are from
the $F\pi$-2~\cite{Horn:2006tm} and $\pi$-CT~\cite{Horn:2007ug}
experiments at JLAB. The numbers displayed in the plots are the average $(Q^2,W)$ values. 
The dashed curves correspond to the exchange of the $\pi$-Regge trajectory alone.
The dash-dotted curves are obtained with the on-mass-shell form
factors in the nucleon-pole contribution and exchange of the
$\rho(770)/a_2(1320)$-trajectory. The solid curves describe the model results with the
resonance contributions. The data points in each $(Q^2,W)$ bin correspond 
to slightly different values of $Q^2$ and $W$ for the various $-t$ bins.
The calculations are performed for values of $Q^2$ and $W$ corresponding to 
the first $-t$ bin. The histograms for $d\sigma_{\rm T}/dt$ are the results 
from~\cite{Kaskulov:2008xc}. 
\vspace{-0.7cm}
}
\end{center}
\end{figure*}

\begin{figure*}
\begin{center}
\includegraphics[clip=true,width=2\columnwidth,angle=0.]{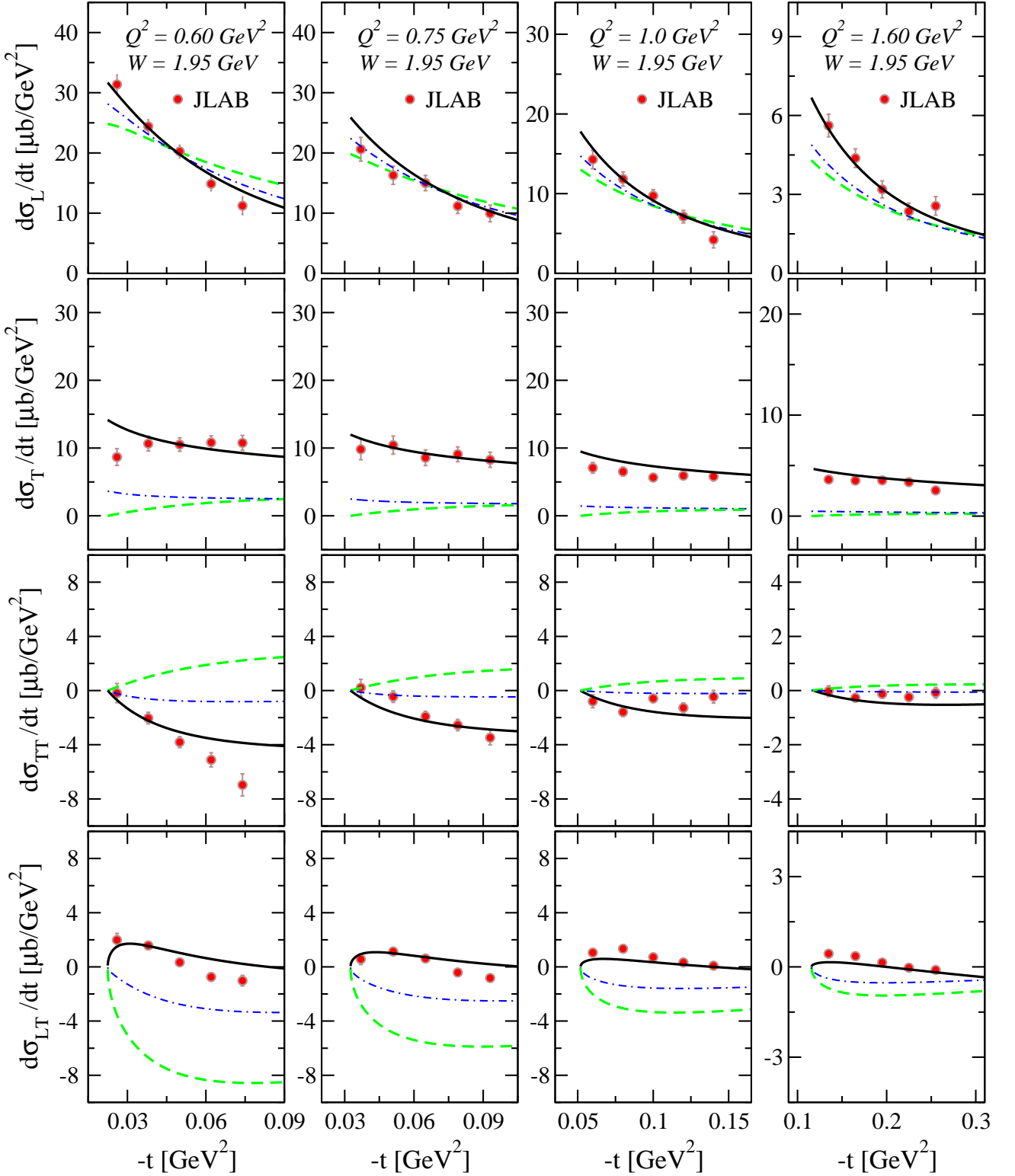}
\caption{\label{Tadv} 
\small (Color online) $-t$ dependence of \textsc{l/t} partial differential
cross sections in exclusive reaction $p(\gamma^*,\pi^+)n$. The experimental
data are from the $F\pi$-1 experiment at JLAB~\cite{Tadevosyan:2007yd}. 
The notations for the curves are the same as in Figure~\ref{Horn1}. 
\vspace{0.7cm}
}
\end{center}
\end{figure*}

\begin{figure*}[t]
\begin{center}
\includegraphics[clip=true,width=2.05\columnwidth,angle=0.]{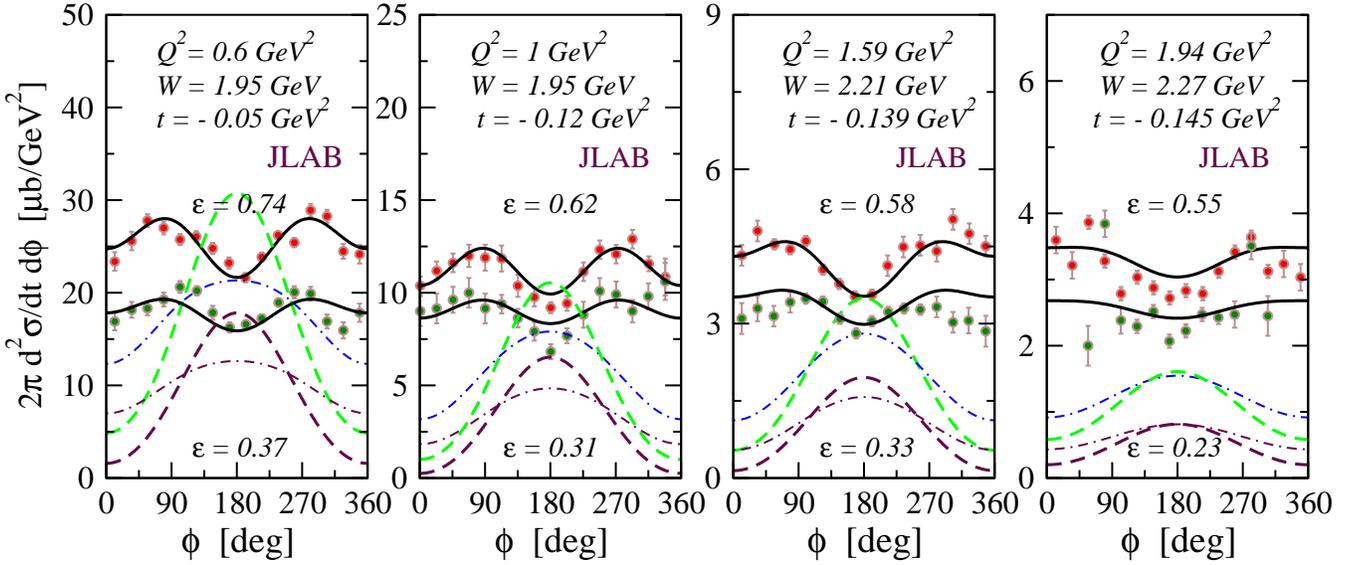}
\caption{\label{PhiDistr} 
\small (Color online) $\phi$ dependence of the 
differential cross sections $2\pi\, d^2\sigma/dtd\phi$ in exclusive reaction
$p(\gamma^*,\pi^+)n$ at fixed $(-t,Q^2,W)$ and two (high and low)
values of $\varepsilon$. The notations for the curves are the same as 
in Figure~\ref{Horn1}. The upper set of solid, dashed and dash-dotted 
curves belong to the higher value of $\varepsilon$. The experimental 
data are from Refs.~\cite{Tadevosyan:2007yd,Horn:2006tm,Horn:2007ug}.
\vspace{-0.cm}
}
\end{center}
\end{figure*}

\section{JLAB $F\pi$-1, $F\pi$-2 and $\pi$-CT data}

In this section we study the \rp effects  in partial $\pi^+$ electroproduction 
cross sections measured at JLAB. We compare the  model  results  with  the  
differential cross sections in the $p(\gamma^*,\pi^+)n$ reaction 
from the $F\pi$-1~\cite{Tadevosyan:2007yd}, $F\pi$-2~\cite{Horn:2006tm}    
and $\pi$-CT~\cite{Horn:2007ug} experiments. 
At JLAB the reaction $n(\gamma^*,\pi^-)p$ has been also measured
off the deuteron target and $\pi^-$ data will be soon
reported~\cite{HuberPriv}.

In Figure~\ref{Horn1} we show our results for the $p(\gamma^*,\pi^+)n$ 
reaction together with the high-$Q^2$ data 
from~\cite{Horn:2006tm,Horn:2007ug}. 
The data points in each $(Q^2,W)$ bin correspond to  slightly different 
values of $Q^2$ and $W$ for the various $-t$ bins.         The numbers 
displayed in the plots are the average $(Q^2,W)$ values. For simplicity 
we perform the calculations for values of  $(Q^2,W)$  corresponding to 
the first $-t$ bin. A proper binning of the curves does not change 
much the results~\cite{Kaskulov:2008xc}.

At first,  we  consider  again  the  reggeized  $\pi$-exchange only. 
The value of the cut-off in the pion from factor is $\Lambda_{\gamma\pi\pi}^2=0.46$~GeV$^2$. 
In  Figure~\ref{Horn1} the dashed curves  describe this contribution. 
As one can see, the steep fall of   $d\sigma_{\rm L}/dt$  away from forward angles comes 
entirely  from  the  rapidly decreasing $\pi$-exchange amplitude. The 
$\pi$-exchange practically saturates the longitudinal 
response $d\sigma_{\rm L}/dt$. At these  values of $Q^2>1.5$~GeV$^2$ the contribution 
of the $\pi$-reggeon exchange to  the transverse cross section $d\sigma_{\rm T}/dt$ 
is already marginal. The $\pi$-exchange  is effective in the interference cross 
sections. 
However, experimentally the cross section 
$d\sigma_{\rm  TT}/dt$ is negative and $d\sigma_{\rm LT}/dt$ is positive. 
The exchange of $\pi$ contributes here just with opposite signs.

We gauge the $\pi$-exchange by adding the nucleon-pole term.   
The exchanges of $\rho(770)/a_2(1320)$ and  $a_1(1260)$ Regge trajectories 
are also added. This result corresponds 
to the dash-dotted curves in  Figure~\ref{Horn1}.   As one can see, the 
longitudinal cross sections $d\sigma_{\rm L}/dt$ is  barely  changed.   
The transverse cross section $d\sigma_{\rm T}/dt$ is negligibly small but 
finite at forward angles. The sign of the  interference cross section 
$d\sigma_{\rm TT}/dt$  now respects the experimental data but the 
magnitude of the cross section  is compatible  with zero. The cross section 
$d\sigma_{\rm LT}/dt$ has been increased by about a factor of two  but is 
still largely negative.

\begin{figure*}
\begin{center}
\includegraphics[clip=true,width=2\columnwidth,angle=0.]{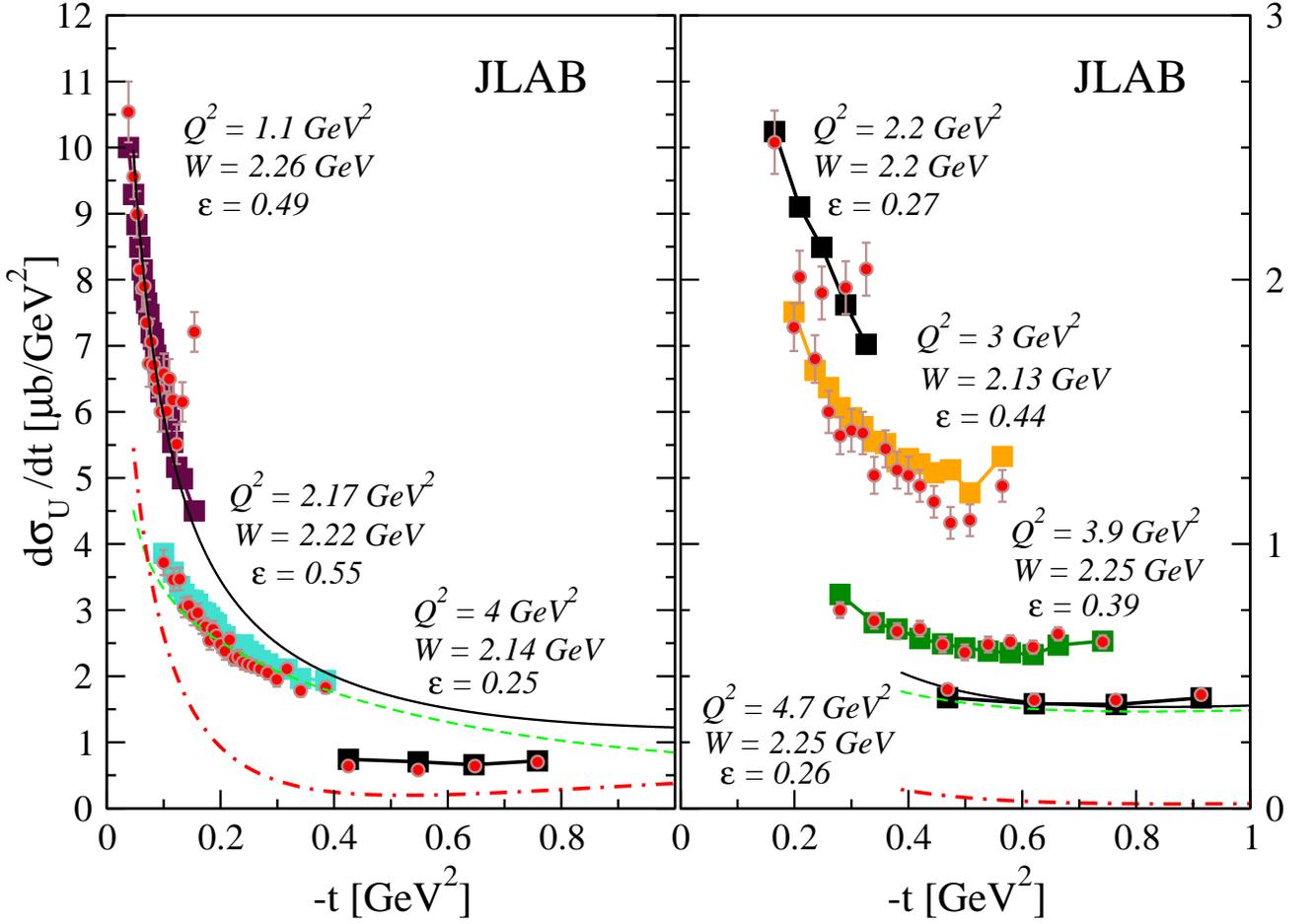}
\caption{\label{Dutta1} (Color online) The differential cross sections
$d\sigma_{\rm U}/dt = d\sigma_{\rm T}/dt + \varepsilon d\sigma_{\rm
L}/dt$ in exclusive reaction $p(\gamma^*,\pi^+)n$
in the kinematics of the $\pi$-CT experiment at JLAB~\cite{Dutta}.
The square symbols connected by solid lines describe the
model results. The discontinuities in the curves result from
the different values of $(Q^2,W,\varepsilon)$ for the various $-t$ bins. 
The dash-dotted and dashed curves describe the contributions of the longitudinal
$\varepsilon d\sigma_{\rm L}$ and transverse $d\sigma_{\rm T}$
cross sections, respectively, to the total unseparated cross sections (solid curves)
for the the lowest and highest average values of $Q^2=1.1$~GeV$^2$ and $Q^2=4.7$~GeV$^2$.}
\vspace{-0.0cm}
\end{center}
\end{figure*}

In the last step we use the \rp transition form factor to include  the 
effect of resonances. The solid curves in Figure~\ref{Horn1} correspond 
to this description. As one can see, all the cross sections are now very 
well described.  Furthermore, the magnitude of  $d\sigma_{\rm T}/dt$  is
strongly  correlated  with the  sign and  magnitude of the interference 
cross  sections  $d\sigma_{\rm TT}/dt$  and  $d\sigma_{\rm LT}/dt$.   The 
description of $d\sigma_{\rm T}/dt$ translates at once into a remarkable 
description  of  both  interference  cross  sections. 
For instance, $d\sigma_{\rm LT}/dt$ changes sign to positive 
and $d\sigma_{\rm TT}/dt$ gets large and negative.
The contribution of resonances to the longitudinal cross section 
$d\sigma_{\rm L}/dt$ is  sizable at forward  angles where the pion form factor
is extracted~\cite{Huber:2008id} and increases with increasing 
value of $Q^2$. However, the effect is particularly pronounced in 
$d\sigma_{\rm T}/dt$ and interference cross sections $d\sigma_{\rm TT}/dt$  
and  $d\sigma_{\rm LT}/dt$. For instance, at $Q^2=2.45$~GeV$^2$ and 
$Q^2=3.91$~GeV$^2$ $d\sigma_{\rm T}/dt$  has been  increased by about 
two orders of magnitude.

The histograms in Figure~\ref{Horn1} for $d\sigma_{\rm T}/dt$ are 
from~\cite{Kaskulov:2008xc}. The assumption used in
Ref.~\cite{Kaskulov:2008xc} for $\gamma^* p \to \pi^+ n$ is that at the  
invariant masses reached at JLAB nucleon resonances 
can contribute to the $1\pi$ channel as well. Then similar to the use of 
Regge trajectories in the $t$-channel that takes higher meson excitations 
into account one has to consider the direct hard interaction of virtual 
photons with partons (DIS) since  DIS involves all possible transitions 
of the nucleon from its ground state to any excited state~\cite{Close:2001ha}.
Modeling the resonance contributions by DIS like processes, followed by
hadronization into the $\pi^+n$ channel, result in histograms 
shown in Figure~\ref{Horn1}. Our present treatment of 
resonance contributions produces a result which is very close to that 
obtained in our previous work~\cite{Kaskulov:2008xc}. However,
the present approach goes beyond the two-component  hadron-parton model of 
Ref.~\cite{Kaskulov:2008xc} and allows to study the interference and 
non-$\pi$-pole background effects on the amplitude level.

The transverse $d\sigma_{\rm T}/dt$ component is insensitive to the variation 
of the cut-off in the pion form factor $\Lambda_{\gamma\pi\pi}$. On the
contrary, the magnitude of $d\sigma_{\rm L}/dt$ is driven by this parameter. 
In Figure~\ref{Tadv} we compare the model results with data measured at lower
values of $(Q^2,W)$ in the $F\pi$-1 experiment~\cite{Tadevosyan:2007yd}.
The calculations are performed for values of  $(Q^2,W)$  corresponding to 
the first $-t$ bin~\cite{Tadevosyan:2007yd,Blok:2008jy}.
The notations for the curves are the same as in Figure~\ref{Horn1}.
In these calculations we used the value of 
$\Lambda_{\gamma\pi\pi}^2=0.4$~GeV$^2$. Also here we find a pronounced
resonance contribution in  $d\sigma_{\rm T}/dt$, $d\sigma_{\rm TT}/dt$  
and  $d\sigma_{\rm LT}/dt$. The slope and magnitude of $d\sigma_{\rm L}/dt$ at 
forward angles are also affected by the resonances.

An extraction of \textsc{l/t} partial differential cross sections  
requires besides the Rosenbluth separation a fit of different harmonics 
in the azimuthal $\phi$-angle  distribution of the measured unseparated 
double differential cross sections. In the actual experiment one measures 
$d^2\sigma/dtd\phi$ for two different $\varepsilon$ bins. In 
Figure~\ref{PhiDistr} we show the $\phi$ dependence of 
$2\pi d^2\sigma/dtd\phi$ in the reaction $p(\gamma^*,\pi^+)n$ at fixed 
$-t$ and two (high and low) values of $\varepsilon$. This is a 
representative example of $\phi$-dependent exclusive cross sections. 
In Figure~\ref{PhiDistr} the solid curves are the model results and 
experimental data are from~\cite{Tadevosyan:2007yd,Horn:2006tm,Horn:2007ug}.
As in Figures~\ref{Horn1} the dashed curves correspond to the contribution
of the $\pi$-exchange and dash-dotted curves do not account for the
resonances. The upper set of solid, dashed and dash-dotted curves belongs 
to the higher value of $\varepsilon$.

\begin{figure*}[t]
\includegraphics[clip=true,width=1.6 \columnwidth,angle=0.]
{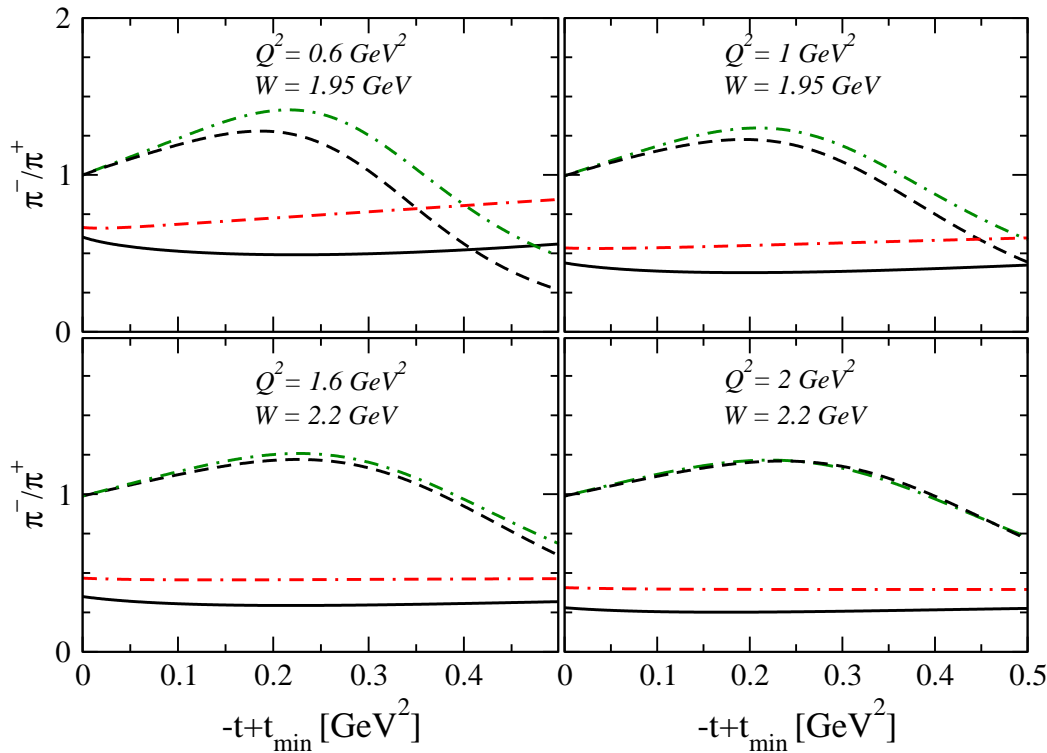}
\caption{\label{RatioPiPPiM} 
(Color online) $-t+t_{min}$ dependence of the ratio of longitudinal 
$R_{\rm L} = d\sigma_{\rm L}^{\pi^-}/d\sigma_{\rm L}^{\pi^+}$ (dashed curves) 
and transverse $R_{\rm T} = d\sigma_{\rm T}^{\pi^-}/d\sigma_{\rm T}^{\pi^+}$  
(solid curves) differential cross sections, 
$n(\gamma^*,\pi^-)p/p(\gamma^*,\pi^+)n$,
for different values of $(Q^2,W)$ at JLAB@5.
The dash-dash-dotted and the dash-dotted curves 
describe $R_{\rm T}$ and $R_{\rm L}$, respectively, without 
$\rho$-reggeon exchange. \vspace{-0.cm}
}
\end{figure*}

\begin{figure*}
\begin{center}
\includegraphics[clip=true,width=2 \columnwidth,angle=0.]
{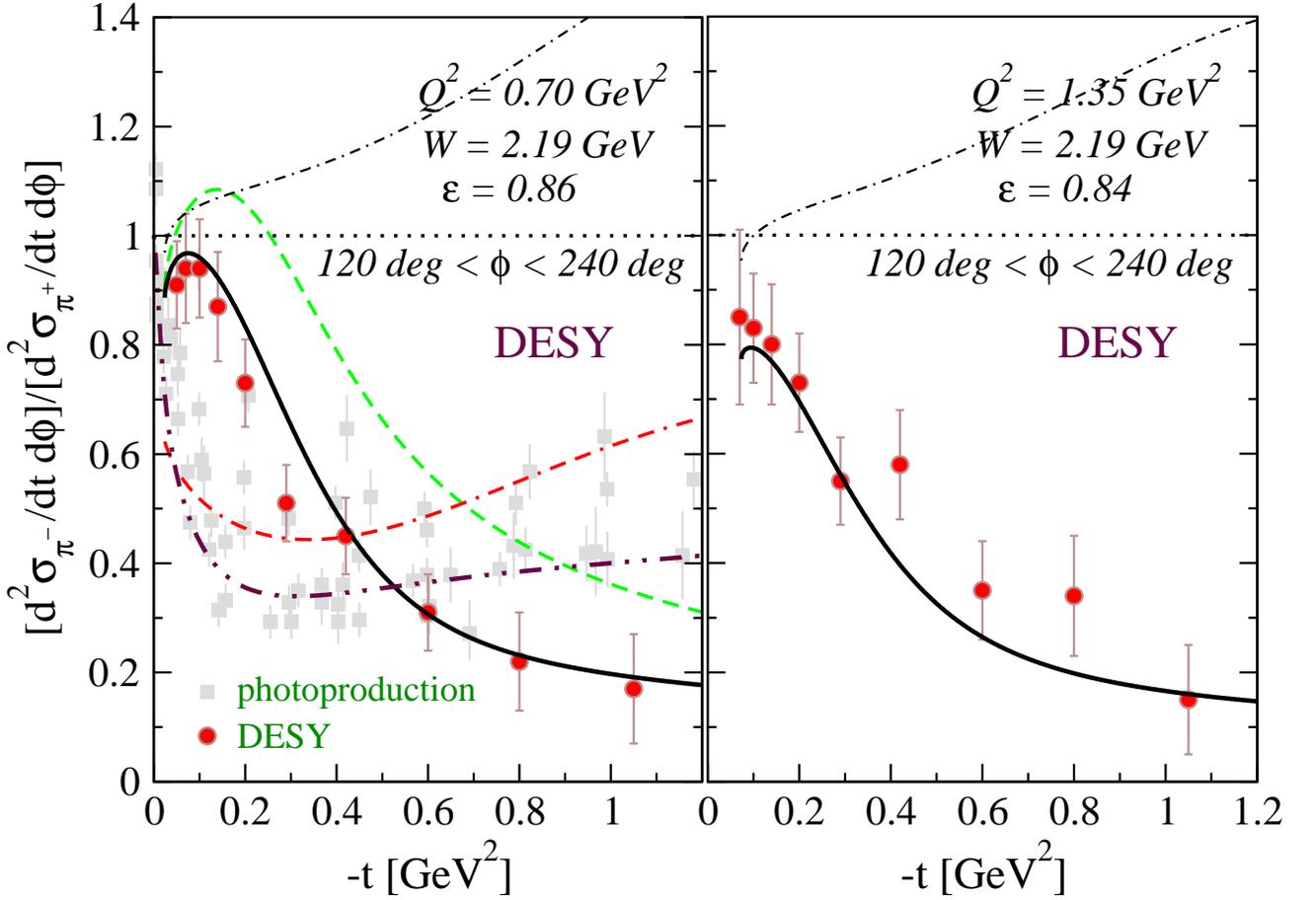}
\caption{\label{RatioPiPPiMDESY} 
\small (Color online)
$-t$ dependence of the $\pi^-/\pi^+$ ratio
of double differential cross sections $d\sigma^2/dtd\phi$ for the value of $W=2.19$~GeV
and at values of $Q^2=0.7$~GeV$^2$, $\varepsilon=0.86$ (left panel) and
$Q^2=1.35$~GeV$^2$, $\varepsilon=0.84$ (right
panel). The experimental data are from Ref.~\cite{DESY4}. The cross sections are integrated in the range of
 $120^{\circ} < \phi < 240^{\circ}$ out-of-plane angles. The solid curves
are the model results. 
The dash-dotted curves describe the results without the resonance contributions.
In the left panel the dashed curve is the results without the exchange
of the $\rho(770)/a_2(1320)$-trajectory. The squares are the compilation of
experimental data for the ratio of $\pi^-/\pi^+$ in photoproduction~\cite{Quinn:1979zp}.
The dot-dot-dashed curve corresponds to the ratio of cross
sections in photoproduction at $Q^2=0$ and $E_{\gamma}=16$~GeV.
The ratio of transverse cross sections at value of
$Q^2=0.7$~GeV$^2$ is given by the dash-dash-dotted curve.
\vspace{-0.0cm}
}
\end{center}
\end{figure*}

\begin{figure*}[t]
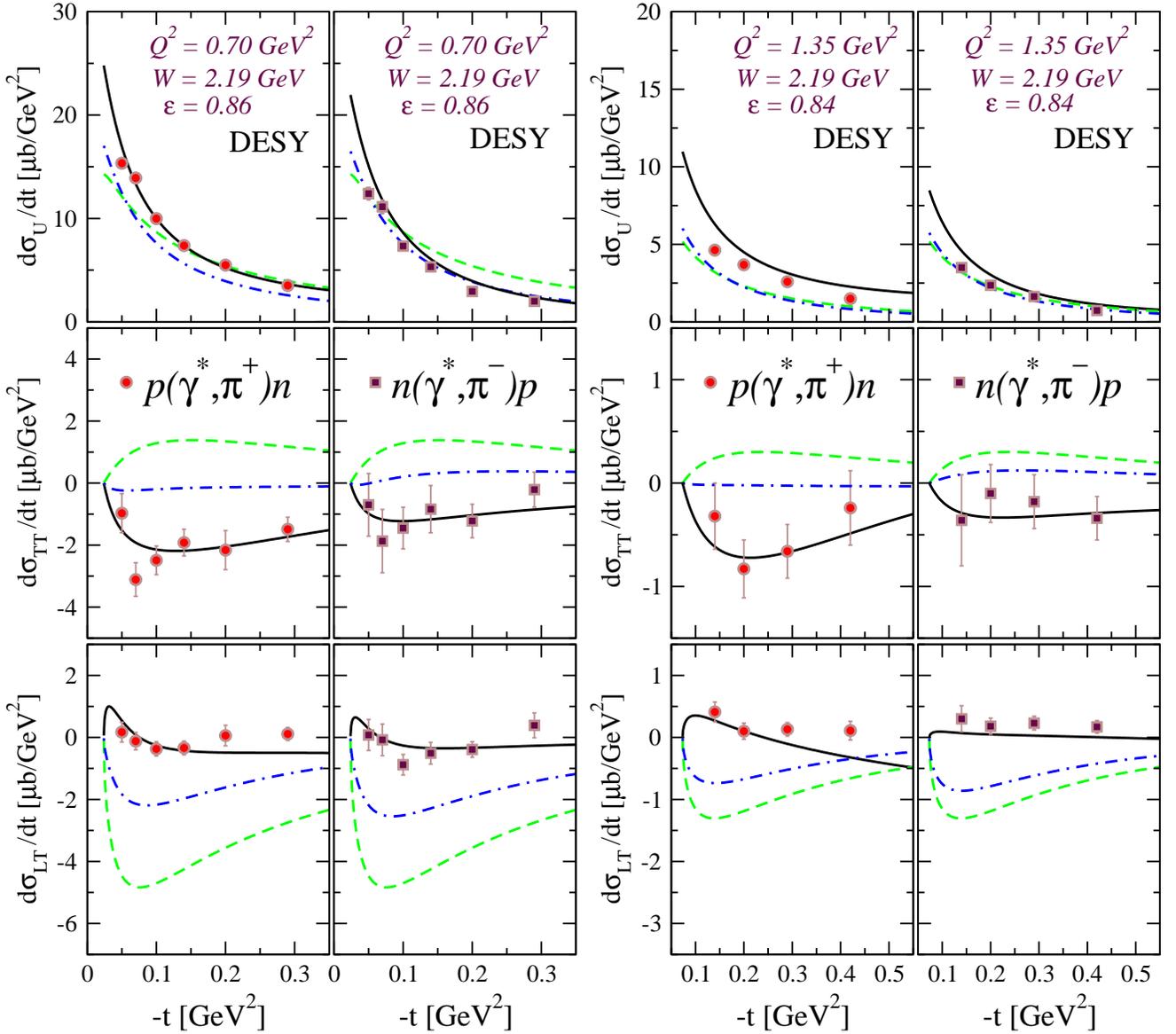

\begin{center}
\includegraphics[clip=true,width=1\columnwidth,angle=0.]{eFig12a.eps}
\includegraphics[clip=true,width=1\columnwidth,angle=0.]{eFig12b.eps}
\caption{\label{DESYPiPPiM} 
\small (Color online) The differential cross sections 
$d\sigma_{\rm U}/dt=d\sigma_{\rm T}/dt + \varepsilon d\sigma_{\rm T}/dt$ (top),
$d\sigma_{\rm TT}/dt$ (middle) and $d\sigma_{\rm LT}/dt$ (bottom)
in exclusive reactions $p(\gamma^*,\pi^+)n$ (left panels) and
$n(\gamma^*,\pi^-)p$ (right panels) in the
kinematics of DESY experiments for the average values of
$Q^2=0.7$~GeV$^2$,
$W=2.19$~GeV, $\varepsilon = 0.86$ and $Q^2=1.35$~GeV$^2$,
$W=2.19$~GeV, $\varepsilon = 0.84$. The notations for the curves are the same
as in Figure~\ref{Horn1}.
The experimental data are from Ref.~\cite{DESY1,DESY2}.
\vspace{-0.0cm}
}
\end{center}
\end{figure*}

\begin{figure}[t]
\begin{center}
\includegraphics[clip=true,width=1\columnwidth,angle=0.]{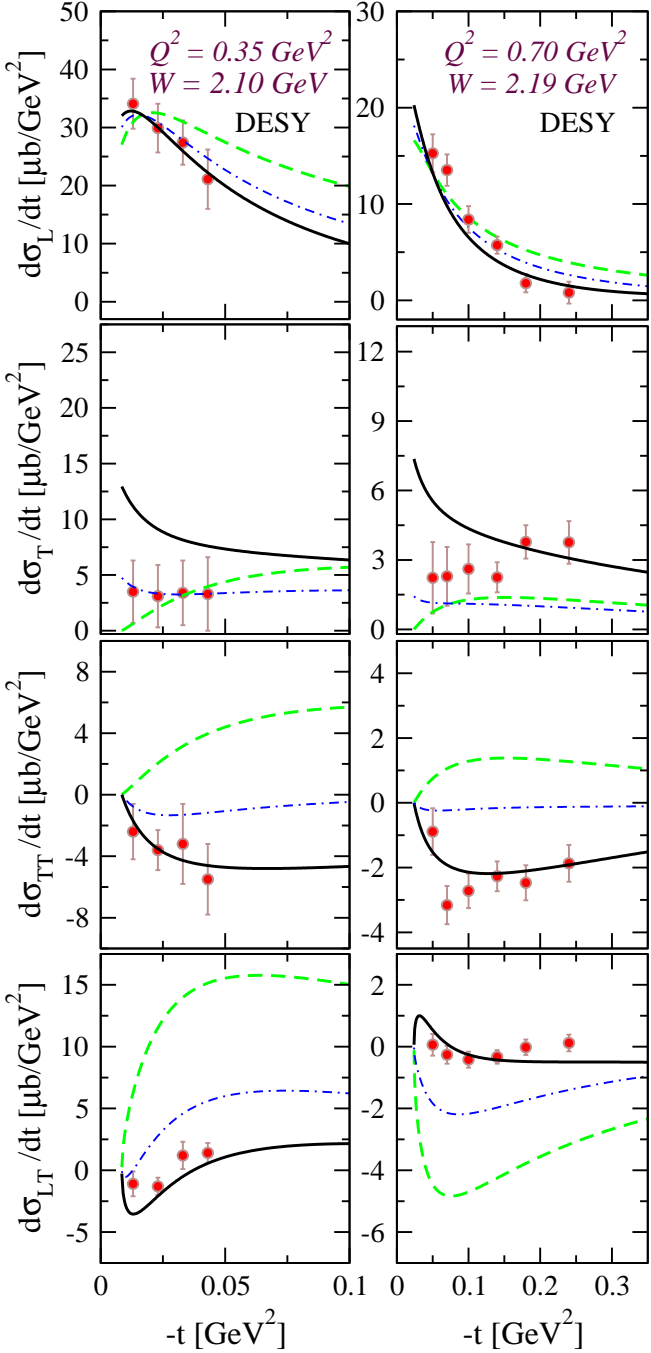}
\caption{\label{DESYPiP} 
\small (Color online) The \textsc{l/t} partial differential cross sections
in the reaction $p(\gamma^*,\pi^+)n$
in the kinematics of DESY experiments. The notations for the curves are the same as in
Figure~\ref{Horn1}. The experimental data for values of $Q^2=0.35$~GeV$^2$ are from
Ref.~\cite{Ackermann:1977rp} and for $Q^2=0.7$~GeV$^2$ are from Ref.~\cite{DESY4}.
\vspace{-0.5cm}
}
\end{center}
\end{figure}

In Figure~\ref{Dutta1} we confront the result of our calculations
with the new JLAB $p(\gamma^*,\pi^+)n$  data~\cite{Dutta}
for unseparated cross sections $d\sigma_{\rm U}/dt$, see Eq.~(\ref{CSU}),
at values of $W \simeq 2.2\div 2.4$~GeV and for different values of
$(Q^2,\varepsilon)$ bins. The square symbols connected by solid lines describe the
model results. The discontinuities in the curves result from
the different values of $(Q^2,W,\varepsilon)$ for the various $-t$ bins. 
The data are very well reproduced by the
present model in the measured $Q^2$ range from $Q^2 \simeq 1$~GeV$^2$
up to $5$~GeV$^2$.  
In Figure~\ref{Dutta1} 
we also show the contributions of the longitudinal
$\varepsilon d\sigma_{\rm L}$ (dash-dotted curves) and transverse $d\sigma_{\rm T}$
(dashed curves) cross sections to the total unseparated cross sections (solid curves)
for the lowest and highest average values of $Q^2=1.1$~GeV$^2$ and $Q^2=4.7$~GeV$^2$.
The cross sections at high values of $Q^2$ are flat and
totally transverse. At forward angles a strong peaking of the cross
section at $Q^2=1.1$~GeV$^2$ comes from the large
longitudinal component in this case. The off-forward region is
transverse. This behavior agrees with the results from~\cite{Kaskulov:2008xc}.
As we shall see, the same behavior is observed in the DIS
regime at HERMES~\cite{:2007an} where the value of $W$ is
higher. At HERMES, because of the Regge shrinkage
of the $\pi$-reggeon exchange and smaller transverse component, the forward peak
just has a steeper $-t$-dependence~\cite{Kaskulov:2009gp}.

We now turn to $\pi^-$ production at JLAB. In~\cite{Kaskulov:2008xc} the transverse 
response $d\sigma_{\rm T}/dt$ in the exclusive reaction $n(\gamma^*,\pi^-)p$ 
was found to be smaller than in the  reaction $p(\gamma^*,\pi^+)n$. 
The present results for the $\pi^-$ channel are parameter free. 
The $u$-channel transition form factors, Eq.~(\ref{F1_u_res}), 
entering $n(\gamma^*,\pi^-)p$  have different behavior, since, contrary to 
the $s$-channel form factors which depend on $(s,Q^2) $ now they depend on
$u$ and $Q^2$ with
\be
\label{Uvar}
u=-s+2M_p^2-t+M_{\pi}^2-Q^2.
\ee

Here we calculate the ratio 
of $\pi^-/\pi^+$ partial cross sections which is of present interest in the dedicated
experiments at JLAB~\cite{HuberPriv}.
In Figure~\ref{RatioPiPPiM} we show the results for the ratio of 
longitudinal $R_{\rm L} = d\sigma_{\rm L}^{\pi^-}/d\sigma_{\rm L}^{\pi^+}$ 
(dashed curves) and transverse $R_{\rm T} = d\sigma_{\rm
  T}^{\pi^-}/d\sigma_{\rm T}^{\pi^+}$ (solid curves) cross
sections as a function of $-t+t_{min}$, where $-t_{min}$
denotes the minimum value of $-t$ for a given $Q^2$ and $W$. The curves have been calculated
for the values of $W=1.95$~GeV (top panels) and $W=2.2$~GeV (bottom panels).
The values of $Q^2$ vary from $Q^2=0.6$~GeV$^2$ (left top)
and $Q^2=1$~GeV$^2$ (right top) to $Q^2=1.6$~GeV$^2$ (left bottom) 
and $Q^2=2$~GeV$^2$ (right bottom).
At forward angles the longitudinal ratio $R_{\rm L}$ 
is close to unity and shows a slow increase
followed by a decrease at higher values of $-t$.
On the contrary, the ratio  $R_{\rm T}$ is practically constant.
For $Q^2=0.6$~GeV$^2$ it is around $R_{\rm T}\simeq 0.6$. With increasing
value of $Q^2$ the ratio $R_{\rm T}$ tends to decrease further and at
$Q^2=2$~GeV$^2$ it gets $R_{\rm T}\simeq 0.26$.

An important mechanism which contributes
to the $\pi^-/\pi^+$ asymmetry is an exchange of the $\rho$-trajectory. 
It is destructive in $\pi^-$ and constructive in $\pi^+$
channels, respectively.
In Figure~\ref{RatioPiPPiM} the dash-dotted (longitudinal ratio) and
dash-dash-dotted curves (transverse ratio) 
correspond to the results without the exchange of $\rho$.
We conclude that $d\sigma_{\rm T}/dt$  in the $\pi^-$ channel is much smaller than in 
the $\pi^+$ production. Also the interference cross sections follow this 
behavior since smaller transverse strength
translates into smaller $d\sigma_{\rm TT}/dt$  
and  $d\sigma_{\rm LT}/dt$.

\section{$p(\gamma^*,\pi^+)n$ and $n(\gamma^*,\pi^-)p$ at DESY}

The early DESY data~\cite{DESY1,DESY2,DESY3,DESY4}  and~\cite{Ackermann:1977rp}
provide an access to the $p(\gamma^*,\pi^+)n$ and
$n(\gamma^*,\pi^-)p$ reactions cross sections in essentially the same $(Q^2,W)$ 
region as at JLAB.
For the proper comparison with data some  
differences in the conventions used by the two different 
groups~\cite{DESY1,DESY2,DESY3,DESY4} 
and~\cite{Ackermann:1977rp}
have to be taken
into account. The $\phi$ convention used here 
follows Refs.~\cite{DESY1,DESY2,DESY3,DESY4}. In~\cite{Ackermann:1977rp} the
azimuthal angle is related to that in~\cite{DESY1,DESY2,DESY3,DESY4} by
$\phi \to \pi-\phi$. The latter results in different signs of the
measured interference cross section $d\sigma_{\rm LT}/dt$.

In Figure~\ref{RatioPiPPiMDESY} we show the results and the 
experimental data~\cite{DESY4} for the ratio of exclusive 
$\pi^-/\pi^+$  double differential cross sections 
$d\sigma^2/dtd\phi$  at the average value of $W=2.19$~GeV and
for the average values of $Q^2=0.7$~GeV$^2$, $\varepsilon = 0.86$ 
(left panel) and $Q^2=1.35$~GeV$^2$, $\varepsilon = 0.84$ (right
panel). The cross sections are integrated in the
 $120^{\circ} < \phi < 240^{\circ}$ azimuthal degree range~\cite{DESY4}. 
The solid curves are the model results. Since, the parameters of the 
model are constrained using the JLAB $p(\gamma^*,\pi^+)n$ data only, 
this agreement with data for the ratio 
$p(\gamma^*,\pi^+)n/n(\gamma^*,\pi^-)p$ of cross sections is indeed
remarkable, see the discussion around Eq.~(\ref{Uvar}). In 
Figure~\ref{RatioPiPPiMDESY} (left panel) the dashed curves are 
the results without the exchange of the $\rho(770)/a_2(1320)$-Regge trajectory. 
It is seen that the $\pi^-/\pi^+$ ratio is indeed sensitive to the 
$\rho$-exchange amplitude. In the left panel we also show the compilation of
experimental data for the ratio of $\pi^-/\pi^+$ photoproduction cross
sections at high energies~\cite{Quinn:1979zp}. For comparison we also show 
the ratio of only the transverse cross sections in electroproduction at 
$Q^2=0.7$~GeV$^2$ (dash-dash-dotted curve).

The dot-dot-dashed curve in Figure~\ref{RatioPiPPiMDESY} (left panel) corresponds to our results
for the ratio of photoproduction ($Q^2=0$) cross sections at
$E_{\gamma}=16$~GeV in the laboratory.
The $\pi^-/\pi^+$ asymmetry seen in the photoproduction 
results mainly from the contribution of the $\rho$-Regge trajectory. However, in
electroproduction the $\pi^-/\pi^+$ asymmetry is driven by the resonance
contributions through the different $(Q^2,s(u))$ dependence of the transition form factors,
Eqs.~(\ref{F1_s_res}) and (\ref{F1_u_res}), 
in the $\pi^+$ and $\pi^-$ channels. For instance, the dash-dotted curves in 
Figure~\ref{RatioPiPPiMDESY} do not account for the contributions of resonances;
the $\pi^-/\pi^+$ ratio is bigger than unity.  In the right panel we show
the results for the values of $Q^2=1.35$~GeV$^2$ and $\varepsilon = 0.84$. 

Our dash-dotted curves which describe the
Regge model without the \rp 
effects are at variance with the results reported in
Ref.~\cite{Vanderhaeghen:1997ts} where the gauge invariant Regge model with
the nucleon-pole contribution has been shown to be in remarkable agreement with
the $\pi^-/\pi^+$ electroproduction ratio. This is surprising since the
model of~\cite{Vanderhaeghen:1997ts} is not compatible with the JLAB
\textsc{l/t} data in the same $(Q^2,W)$ region~\cite{Blok:2008jy}.

\begin{figure*}[t]
\begin{center}
\includegraphics[clip=true,width=2.07\columnwidth,angle=0.]
{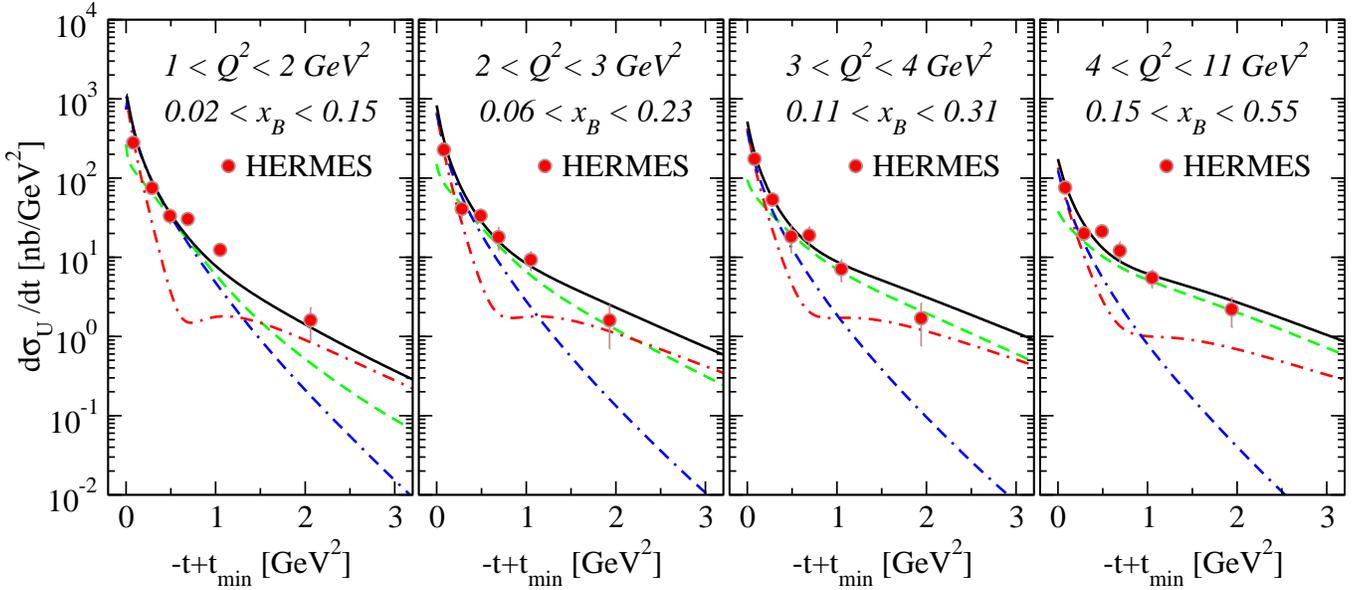}
\caption{\label{EffHermes} (Color online) $-t+t_{min}$ dependence of the 
differential cross section $d\sigma_U/dt = d\sigma_{\rm T}/dt + \epsilon d\sigma_{\rm
L}/dt$ in exclusive reaction $p(\gamma^*,\pi^+)n$ at HERMES. 
The experimental data are from Ref.~\cite{:2007an}. 
The calculations are performed for the average values of
$(Q^2,x_{\rm B})$ in a given $Q^2$ and Bjorken $x_{\rm B}$ bin.
The solid curves are the full model results.
The dash-dotted curves correspond to the longitudinal $\epsilon d\sigma_{\rm L}/dt$
and the dashed curves to the transverse $d\sigma_{\rm T}/dt$ components of the
cross section.
The dash-dash-dotted curves describe the results without the
resonance/partonic effects.}
\vspace{-0.5cm}
\end{center}
\end{figure*}

\begin{figure}[b]
\begin{center}
\includegraphics[clip=true,width=1.\columnwidth,angle=0.]
{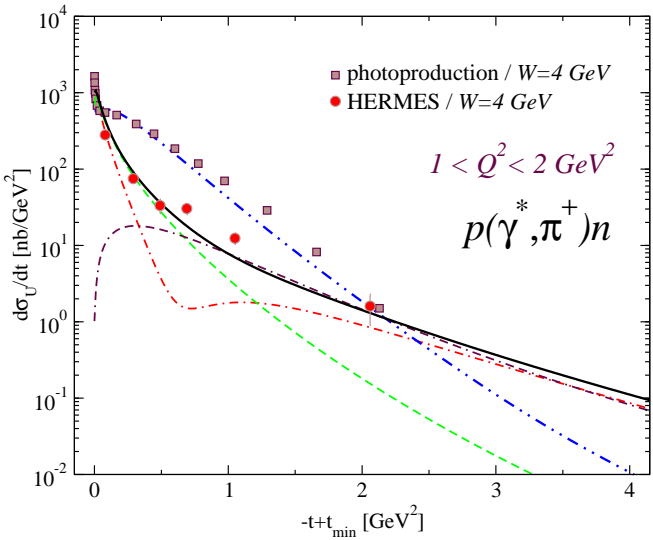}
\caption{\label{EffHermesPiPPiM} (Color online) 
$d\sigma_{\rm U}/dt$ in exclusive reaction $p(\gamma^*,\pi^+)n$ 
for the average values of $Q^2=1.4$~GeV$^2$ and $x_{\rm B}=0.08$.
The HERMES data, the solid and the dash-dotted curves are the same as in the left panel of
Figure~\ref{EffHermes}.
The dashed and dash-dash-dotted curves describe the contributions of the 
$\pi$-exchange and resonance/partonic mechanisms to $\epsilon d\sigma_{\rm
  L}/dt$, respectively. The dot-dot-dashed curve describe
the $\pi^+$ photoproduction for $E_{\gamma}=8$~GeV ($W\simeq 4$~GeV). The
photoproduction data are from~\cite{Boyarski:1967sp}.
}
\vspace{-0.5cm}
\end{center}
\end{figure}

Before drawing definite conclusions concerning these discrepancies, 
we compare in Figure~\ref{DESYPiPPiM} our model results with the measured 
differential cross sections 
$d\sigma_{\rm U}/dt=d\sigma_{\rm T}/dt + \varepsilon d\sigma_{\rm T}/dt$ (top),
$d\sigma_{\rm TT}/dt$ (middle) and $d\sigma_{\rm LT}/dt$ (bottom)
in exclusive reactions $p(\gamma^*,\pi^+)n$ (left panels) and
$n(\gamma^*,\pi^-)p$ (right panels). The experimental data are from
Refs.~\cite{DESY1,DESY2}. The
average values of $(Q^2,W,\varepsilon)$ are the same as in
Figure~\ref{RatioPiPPiMDESY} for the $\pi^-/\pi^+$ ratio. The dashed curves
which refer to the contribution of the $\pi$-exchange are equal in both
channels. The solid curves describe the model results and, as one can see, 
it is not surprising why the $\pi^-/\pi^+$ ratio of double differential cross 
sections, which involves all four \textsc{l/t}  components, is well reproduced. 
A model without the resonance contributions (dash-dotted curves in
Figure~\ref{DESYPiPPiM}) fails in both the $p(\gamma^*,\pi^+)n$ and
$n(\gamma^*,\pi^-)p$ channels and the behavior of the corresponding results
 (dash-dotted curves in
Figure~\ref{RatioPiPPiMDESY}) is expected for the $\pi^-/\pi^+$ ratio.

In Figure~\ref{DESYPiP} (left panel) we compare the results of our
calculations with $\textsc{l/t}$ separated $p(\gamma^*,\pi^+)n$ data 
from~\cite{Ackermann:1977rp} with average values of $Q^2=0.35$~GeV$^2$ 
and $W=2.1$~GeV. These data are kinematically close  to the real 
photon point where one does not expect a variation of the pion form 
factor from its VMD value with $\Lambda_{\gamma\pi\pi}^2=m_{\rho(770)}^2 \simeq
0.59$~GeV$^2$. The 
curves correspond to this choice of $\Lambda_{\gamma\pi\pi}$.  In the 
right panel we also show the $\textsc{l/t}$ separated cross section 
at $Q^2=0.7$~GeV$^2$ and $W=2.19$~GeV. The notations for the curves 
are the same as in Figure~\ref{DESYPiPPiM}. As one can notice, 
because of different conventions for $\phi$, the signs of the 
measured interference cross section $d\sigma_{\rm LT}$ are different 
for the two data sets. 

The interference pattern between the meson-exchange and resonance contributions is
different in the $\pi^+$ and $\pi^-$ channels. For $\pi^-$ production there
are no data for the longitudinal and transverse cross sections. However, we 
conclude that the present model describes rather well the available $\pi^+$  
and $\pi^-$ data for the unseparated cross sections $d\sigma_{\rm U}/dt$
and the interference cross sections $d\sigma_{\rm TT}/dt$ and $d\sigma_{\rm LT}/dt$.
Concerning the $\pi^-/\pi^+$ ratio, we have seen that
the resonance contributions are important and that, contrary to the results of
Ref.~\cite{Vanderhaeghen:1997ts}, the model based on reggeon exchanges is not compatible with
the observed ratio\footnote{Attempting to resolve the latter discrepancy we assumed
that in Figures~2 and~3 of Ref.~\cite{Vanderhaeghen:1997ts} the 
convention of Ref.~\cite{Ackermann:1977rp}  is used for $\phi$. This is fine
for the data set from~\cite{Ackermann:1977rp} in Figures~3 of
Ref.~\cite{Vanderhaeghen:1997ts} but is not correct for the
data set from~\cite{DESY2} presented in
Figures~2 of Ref.~\cite{Vanderhaeghen:1997ts}. Interestingly, then the 
$\pi^-/\pi^+$ ratios of Ref.~\cite{Vanderhaeghen:1997ts} are very well reproduced.}.

\section{Deeply virtual $p(e,e'\pi^+)n$ at Hermes}
\label{DIS_Hermes}

The HERMES data at DESY~\cite{:2007an} in exclusive reaction $p(e,e'\pi^+)n$ 
extend the kinematic region to higher values of $W^2\simeq 16$~GeV$^2$ 
toward the DIS region and higher values of $-t$. At HERMES the kinematic 
requirement $Q^2>1$~GeV$^2$ has been imposed on the scattered electron 
in order to select the hard scattering regime. The resulting range is 
$1<Q^2<11$~GeV$^2$ and $0.02<x_{\rm B}<0.55$ for the Bjorken variable. 
The measured cross sections have been integrated over the azimuthal 
angle $\phi$ and a separation of the transverse and longitudinal parts 
was not feasible. With the $27.6$~GeV HERA beam energy the ratio of 
longitudinal to transverse polarization of the virtual photon $\varepsilon$
is close to unity.

The results for the unseparated differential cross sections 
$d\sigma_{\rm U}/dt=d\sigma_{\rm T}/dt+\varepsilon d\sigma_{\rm L}/dt$ in 
deeply virtual $p(\gamma^*,\pi^+)n$ reaction at HERMES are shown in 
Figure~\ref{EffHermes}. We perform the calculations for the average 
$(Q^2,x)$ values in a given $Q^2$ and $x_{\rm B}$ bin~\cite{AirapetianPriv}.
In Figure~\ref{EffHermes} instead of $t$, the quantity $-t+t_{min}$ is 
again used, where $-t_{min}$ denotes the minimum value of $-t$ for a given 
$Q^2$ and $x_{\rm B}$. The different panels in Figure~\ref{EffHermes} 
correspond to the  different $Q^2$ and $x$ bins. In the calculations we 
use the cut-off $\Lambda_{\gamma\pi\pi}^2=0.46$~GeV$^2$ in the pion form factor,
Eq.~(\ref{PiFF}). This is an optimal value needed for the description of 
high $Q^2$ data at JLAB. In Figure~\ref{EffHermes} the model results which 
include both the meson-exchange and \rp contributions are 
shown by the solid curves. The dash-dotted and 
the dashed curves correspond to the longitudinal $\epsilon d\sigma_{\rm L}/dt$ 
and to the transverse $d\sigma_{\rm T}/dt$  components of the cross section, 
respectively. The dash-dash-dotted curves describe the results without the
\rp contributions.

\begin{figure*}
\begin{center}\includegraphics[clip=true,width=2.0\columnwidth,angle=0.]
{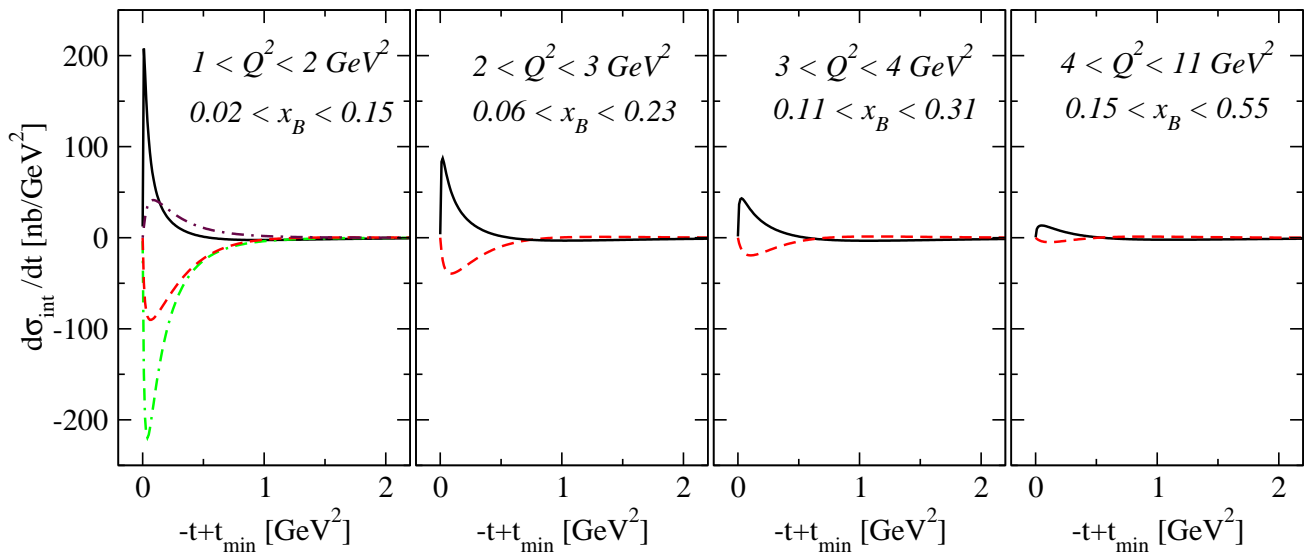}
\caption{\label{EffHermesInt} (Color online)   $-t+t_{min}$ dependence of the 
interference cross sections 
$d\sigma_{\rm TT}$ (dashed curves) and $d\sigma_{\rm LT}$ (solid curves)  at
HERMES. The average values of $Q^2$ and $x_{\rm B}$ used for different
$(Q^2,x_{\rm B})$ bins are the same as in Figure~\ref{EffHermes}.
In the left panel the dash-dotted and dash-dash-dotted curves describe 
the contribution of the $\pi$-reggeon exchange to $d\sigma_{\rm TT}$  and $d\sigma_{\rm LT}$, respectively.}
\vspace{-0.7cm}
\end{center}
\end{figure*}

Interestingly, the physics of $p(\gamma^*,\pi^+)n$ in the DIS region at HERMES 
is essentially the same as at JLAB where the value of $W$ is smaller (2~GeV {\it
  vs.} 4~GeV). Just contrary to the situation in the JLAB experiment the 
longitudinal cross section at HERMES determines the total differential 
cross section at small $-t$.
As at JLAB the transverse cross section at HERMES is
dominated by the \rp mechanism. 
At JLAB the transverse cross section is somewhat larger and at forward angles comparable
with the longitudinal cross section. In deeply virtual production of $\pi^+$ at HERMES the
transverse cross section gets smaller at forward angles 
and the cross section is
dominated by the exchange of Regge trajectories, with $\pi$ being the dominant
trajectory. 
The $\pi$-reggeon exchange contributes mainly to the longitudinal response
$\sigma_{\rm L}$ and at low momentum
transfer $-t$ the variation of the forward cross section with $Q^2$ falls down as
the electromagnetic form factor of the pion
$\sigma_{\rm L} \propto (F_{\gamma\pi\pi} (Q^2))^2$.
In the off-forward region, $-t>1$~GeV$^2$, because of the exponential fall-off of
Regge contributions as a function of $-t$, the
meson-exchange processes are already negligible. 
Above $-t>1$~GeV$^2$ the model cross section
points mainly toward the direct coupling of the virtual photons to
partons. Indeed, this is rather natural, since with 
increasing $-t$ at fixed $Q^2$  smaller distances
can be accessed. This is opposed to $t$-channel meson-exchange 
processes which involve peripheral production of $\pi^+$.

\begin{figure}[t]
\includegraphics[clip=true,width=1\columnwidth,angle=0.]
{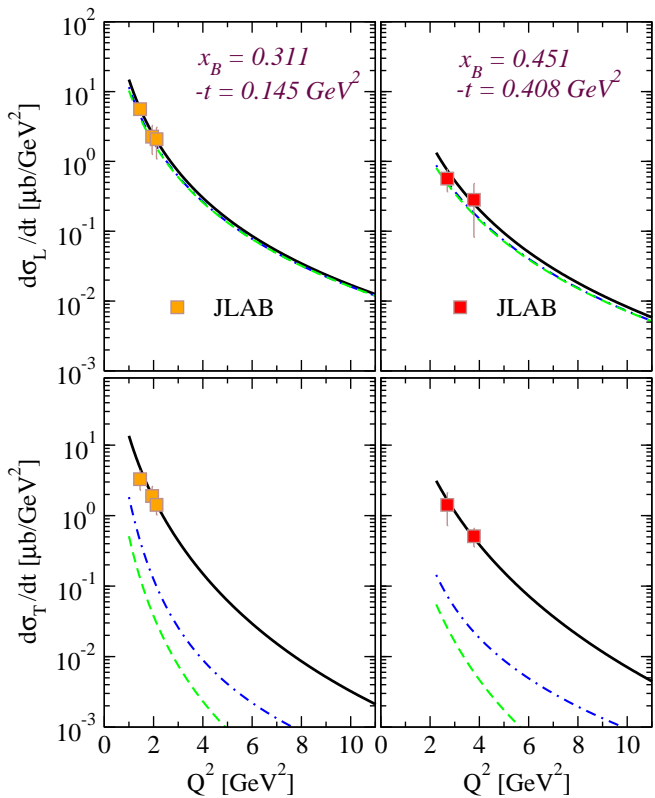}
\caption{\label{EffHornScalingQ2} 
\small (Color online) $Q^2$ dependence of the longitudinal $d\sigma_{\rm
  L}/dt$ (top panels) and transverse $d\sigma_{\rm T}/dt$ (bottom panels) 
cross sections in $p(\gamma^*,\pi^+)n$ reaction at fixed values of
$-t$ and Bjorken $x_{\rm B}$. The solid curves are the model predictions for
the scaling curves. The dashed curves correspond to the contribution 
of the $\pi$-reggeon exchange alone. The dash-dotted curves are the model
results without the contributions of resonances.
The experimental data are from Ref.~\cite{Horn:2007ug}. 
\vspace{-0.0cm}
}
\end{figure}

\begin{figure}[t]
\includegraphics[clip=true,width=1\columnwidth,angle=0.]
{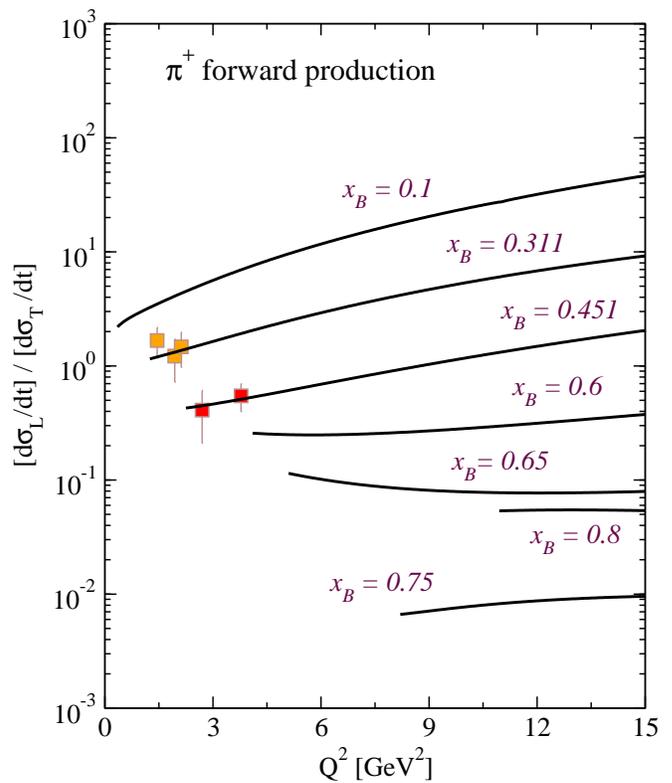}
\caption{\label{xBScalingQ2} 
\small (Color online)  
The $Q^2$ dependence of the ratio of longitudinal $d\sigma_{\rm L}/dt$ to transverse $d\sigma_{\rm
  T}/dt$ differential cross sections in the forward $\pi^+$ production. 
The different curves correspond to different values of Bjorken scaling
variable $x_{\rm B}$.
\vspace{-0.0cm}
}
\end{figure}

\begin{figure*}
\includegraphics[clip=true,width=2\columnwidth,angle=0.]
{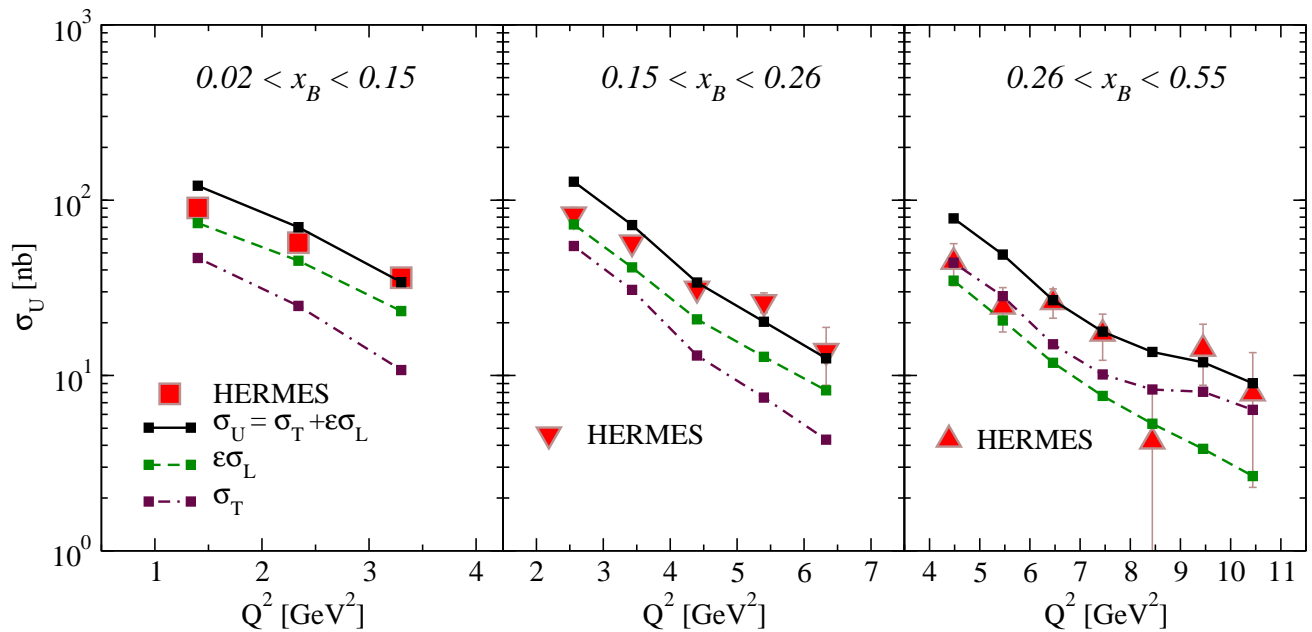}
\caption{\label{HermesQ2dep} (Color online) 
$Q^2$ dependence of the integrated cross sections 
$\sigma_{\rm U}=\sigma_{\rm T}+\epsilon \sigma_{\rm L}$
in exclusive reaction $p(\gamma^*,\pi^+)n$ at HERMES. 
The different panels correspond to different $x_B$ bins. 
The solid curves are the model results. 
The dashed and dash-dotted curves correspond to the longitudinal $\epsilon \sigma_{\rm L}$
and transverse $\sigma_{\rm T}$ components
of the cross section, respectively. The experimental data are from
Ref.~\cite{:2007an}} 
\vspace{-0.cm}
\end{figure*}

Contrary to the two-component model of Ref.~\cite{Kaskulov:2009gp}, 
in the present model there is a sizable longitudinal \rp
component which is effective in the off-forward $\pi^+$ production at low 
values of Bjorken $x_{\rm B}$. For instance, the dip in 
$\varepsilon d\sigma_{\rm L}/dt$ , see the dash-dotted curves
in Figure~\ref{EffHermes}, results from the interference between the 
meson-exchange and the \rp contributions. In Figure~\ref{EffHermesPiPPiM} 
we show the different contributions to $\varepsilon d\sigma_{\rm L}/dt$ 
for the lowest $x_{\rm B}$ bin. The solid and dash-dotted curves in 
Figure~\ref{EffHermesPiPPiM} are the same as in Figure~\ref{EffHermes} (left panel). 
The contribution of the $\pi$-exchange to  $\varepsilon d\sigma_{\rm L}/dt$
is shown by the dashed curve. The latter  falls exponentially down as a function of
$-t$. The dash-dash-dotted curve describe the \rp contribution
to $\varepsilon d\sigma_{\rm L}/dt$. It vanishes at forward angles and by
the destructive interference with the $\pi$-reggeon exchange produces a dip
in $\varepsilon d\sigma_{\rm L}/dt$. With increasing $-t$ the partonic component of 
$\varepsilon d\sigma_{\rm L}/dt$ continues to dominate the longitudinal response.
 
In Figure~\ref{EffHermesPiPPiM} the dot-dot-dashed curve describes the real photon
limit $Q^2=0$ of the model cross section for $E_{\gamma}=8$~GeV ($W\simeq 4$~GeV). The
photoproduction data are from
Ref.~\cite{Boyarski:1967sp}. This is the same $W$ region as in the HERMES
electroproduction  data. The model is in agreement with both the photo- and 
electroproduction data. 

In Figure~\ref{EffHermesInt} we present the results for the interference cross
sections at HERMES energies.  
$d\sigma_{\rm TT}/dt$ and $d\sigma_{\rm LT}/dt$ are sizable, 
and follow a behavior observed already at JLAB energies. For instance,
if the cross section would be dominated by the $\pi$-reggeon exchange then
$d\sigma_{\rm TT}/dt$ would be positive (dash-dotted curves in the left panel) and $d\sigma_{\rm
  LT}/dt$  negative (dash-dash-dotted curves in the left panel). The interference between the
$\pi$-trajectory and \rp contributions change the sign of
both $d\sigma_{\rm TT}/dt$ and $d\sigma_{\rm LT}/dt$. The solid and dashed
curves describe the model results for $d\sigma_{\rm LT}/dt$ and
$d\sigma_{\rm TT}/dt$, respectively.

\section{$Q^2$ dependence of the cross sections}

It has been proposed that the $Q^2$ dependence of $\textsc{l/t}$ separated exclusive
$p(\gamma^*,\pi^+)n$ cross sections may provide a test of the factorization 
theorem~\cite{Collins:1996fb} in the separation of long-distance and 
short-distance physics and the extraction of GPD.  The leading twist GPD 
scenario predicts for $\sigma_{\rm L}\sim 1/Q^6$
and $\sigma_{\rm T}\sim 1/Q^8$. An observation of the $Q^2$ power law scaling
is considered as a model independent test of QCD factorization.

The $Q^2$ behavior of cross sections in exclusive reaction $p(\gamma^*,\pi^+)n$ 
has been studied at JLAB in Ref.~\cite{Horn:2007ug}. 
It was shown that while the scaling laws are
reasonably consistent with the $Q^2$ dependence of the longitudinal $\sigma_{\rm L}$ data,
they fail to describe the $Q^2$ dependence of the transverse $\sigma_{\rm T}$ data.
The $Q^2$ dependence of the $p(\gamma^*,\pi^+)n$ cross section in DIS has been
also studied at HERMES~\cite{AirapetianPriv}. It was found that the $Q^2$
dependence
of the data is in general well described by the calculations from GPD models
which include the power corrections,
see Ref.~\cite{AirapetianPriv} and references therein. However, the magnitude of the theoretical cross
section is underestimated. The Regge model of Ref.~\cite{Laget:2004qu} was shown to be
compatible with, both, $-t$ and $Q^2$ dependencies of the HERMES data.
In the following we check this predicted $\sigma_{\rm L} / \sigma_{\rm T}
\sim Q^2$ scaling within our model calculations.

\begin{figure*}
\includegraphics[clip=true,width=1.7 \columnwidth,angle=0.]{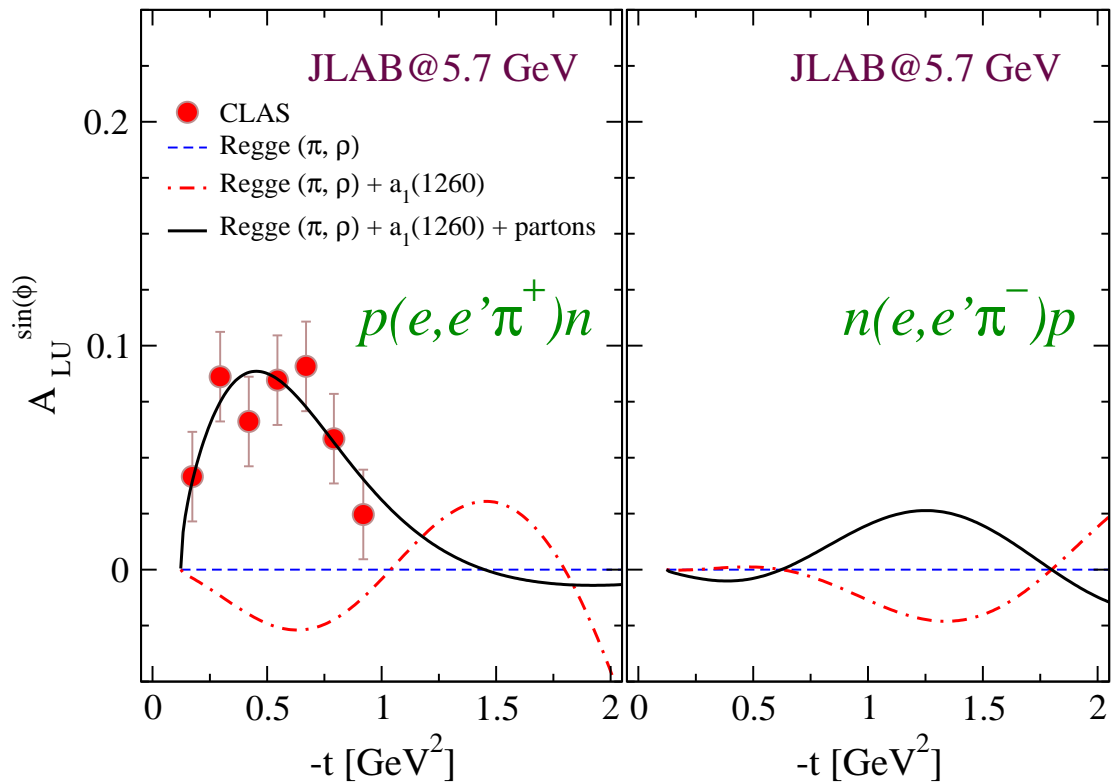}
\caption{\label{BSAAvakian}
\small (Color online) Left panel: The beam spin azimuthal moment
$A^{\sin(\phi)}_{\rm LU}$ in exclusive
reaction $p(\gamma^*,\pi^+)n$ as a function of $-t$. The CLAS/JLAB data~\cite{Avakian:2004dt} 
were collected for hard scattering kinematics with
values of $E_e=5.77$~GeV, $Q^2>1.5$~GeV$^2$ and $W^2>4$~GeV$^2$.  
The dashed curves describe 
the results (the asymmetry is zero) without the resonance contributions 
and neglecting the exchange of unnatural parity  $a_1(1260)$ Regge trajectory.  
The dash-dotted curves correspond
to the addition of the axial-vector $a_1(1260)$-reggeon exchange. 
The solid curves are the model results and account for the
resonance/partonic effects.
Right panel:
The beam spin azimuthal moment $A^{\sin(\phi)}_{\rm LU}$ in exclusive reaction
$n(\gamma^*,\pi^-)p$.
The notations for the curves are the same as in the left panel.
\vspace{-0.0cm}
}
\end{figure*}

\begin{figure*}[t]
\begin{center}
\includegraphics[clip=true,width=1.7 \columnwidth,angle=0.]{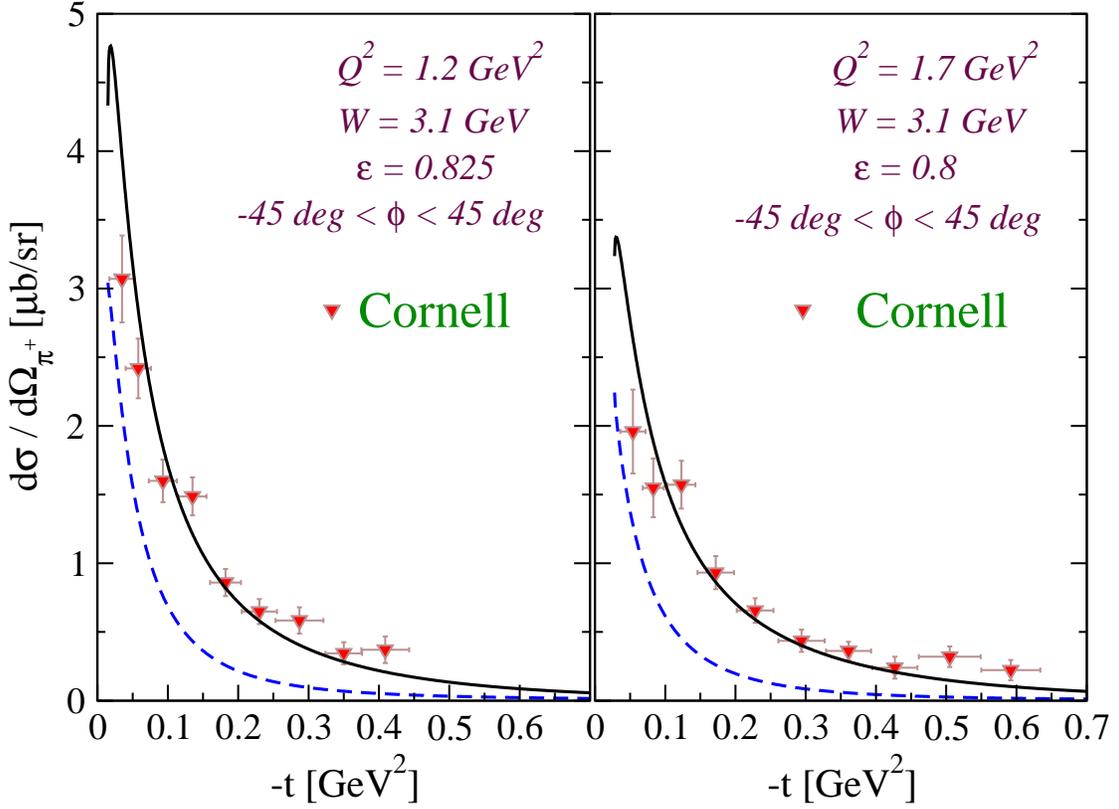}
\caption{\label{Cornell3GeV12and17}
\small (Color online) The differential cross sections for the electroproduction of
$\pi^+$ meson at Cornell. The solid curves describe the model results and
the dashed curves correspond to the results without the resonance/partonic contributions.
The data are from \cite{Cornell_2}. 
The values of $(W,Q^2,\varepsilon)$ are the average ones. The theoretical
cross sections have been integrated over
the range of azimuthal out-of-plane acceptance $-45^{\circ} < \phi < 45^{\circ}$.
\vspace{-0.5cm}
}
\end{center}
\end{figure*}

\subsection{JLAB data}

In Figure~\ref{EffHornScalingQ2} we show our results for the $Q^2$ 
dependence of $p(\gamma^*,\pi^+)n$ reaction cross sections 
$d\sigma_{\rm L}/dt$ and $d\sigma_{\rm T}/dt$ at fixed $-t$ and
Bjorken variable $x_B$. The experimental data are from Ref.~\cite{Horn:2007ug}
and correspond to the forward $\pi^+$ production. The solid curves 
are the model predictions and describe the available data
very well. The dashed curves describe the contribution of the
$\pi$-reggeon exchange to the $Q^2$ scaling curves only. The dash-dotted 
curves are the model results without the resonance contributions. 
The latter effect is again large in the transverse cross section and 
gives only small correction to the longitudinal cross section 
$d\sigma_{\rm L}/dt$. The $Q^2$ dependence of $d\sigma_{\rm L}/dt$ is 
essentially driven by the pion form factor.

The $Q^2$ dependence of the ratio of longitudinal $d\sigma_{\rm L}/dt$ to 
transverse $d\sigma_{\rm  T}/dt$ differential cross sections for the 
forward $\pi^+$ production is shown in Figures~\ref{xBScalingQ2}. 
The different curves correspond to different values of $x_{\rm B}$. 
All the curves start at the value of $W \simeq 1.9$~GeV. For small and 
intermediate values of $x_{\rm B}$ the model results show an increase of 
the ratio $d\sigma_{\rm L}/d\sigma_{\rm T}$ as a function of $Q^2$. Only at 
small values of Bjorken $x_{\rm B}$ the ratio $d\sigma_{\rm L}/d\sigma_{\rm
  T}$ is qualitatively in agreement with the predicted $\sim Q^2$ behavior.  
In the valence quark region above $x_{\rm B} \simeq 0.6$ the cross
section ratio scales and is actually independent of the value of $Q^2$. In
this region the transverse component  $\sigma_{\rm T}$ dominates the $\pi^+$ 
electroproduction cross section. In the experimental determination of the pion
transition form factor from forward $\sigma_{\rm L}$ data one can, therefore,
better isolate the longitudinal response by minimizing the Bjorken $x_{\rm B}$.

\begin{figure}[b]
\begin{center}
\includegraphics[clip=true,width=1.0 \columnwidth,angle=0.]{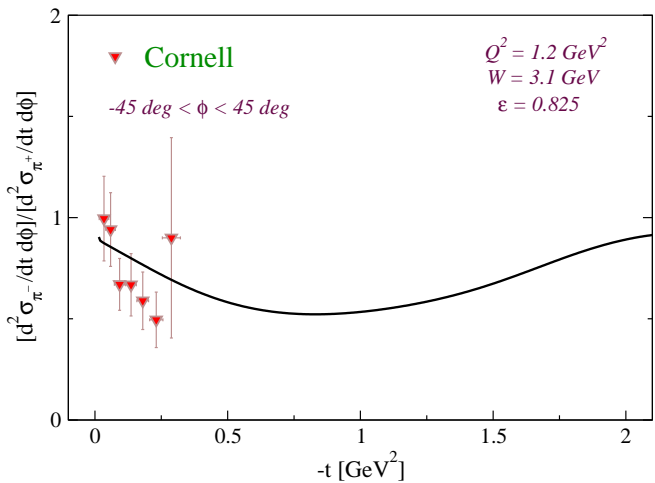}
\caption{\label{CornellRatio3GeV12}
\small (Color online) The $\pi^-/\pi^+$ ratio of differential cross sections for the electroproduction of
$\pi^-$ and $\pi^+$ mesons at Cornell. 
The data are from \cite{Cornell_2}.  The theoretical
cross section has been integrated over
the range of azimuthal out-of-plane acceptance $-45^{\circ} < \phi < 45^{\circ}$.
\vspace{-0.5cm}
}
\end{center}
\end{figure}

\begin{figure*}
\includegraphics[clip=true,width=2 \columnwidth,angle=0.]{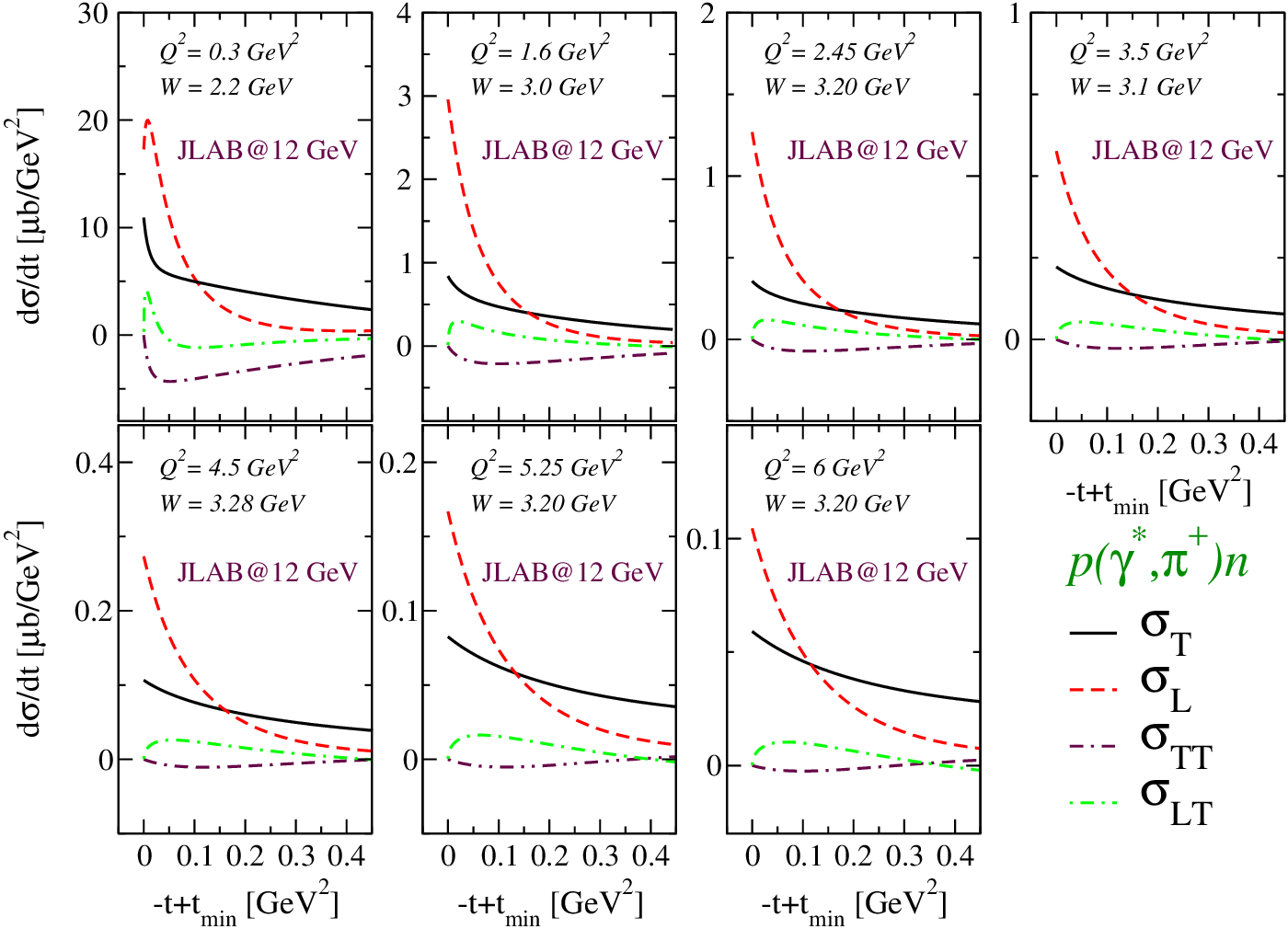}
\caption{\label{JLAB12b} (Color online) $-t+t_{min}$ dependence of 
\textsc{l/t} separated differential cross sections $d\sigma_{\rm T}/dt$
(solid), $d\sigma_{\rm L}/dt$ (dashed), $d\sigma_{\rm LT}/dt$ (dash-dotted)
and  $d\sigma_{\rm TT}/dt$ (dash-dash-dotted) in exclusive reaction
$p(\gamma^*,\pi^+)n$ in the $(Q^2,W)$ kinematics at JLAB@12.
\vspace{-0.2cm}}

\end{figure*}

\begin{figure*}
\includegraphics[clip=true,width=2 \columnwidth,angle=0.]{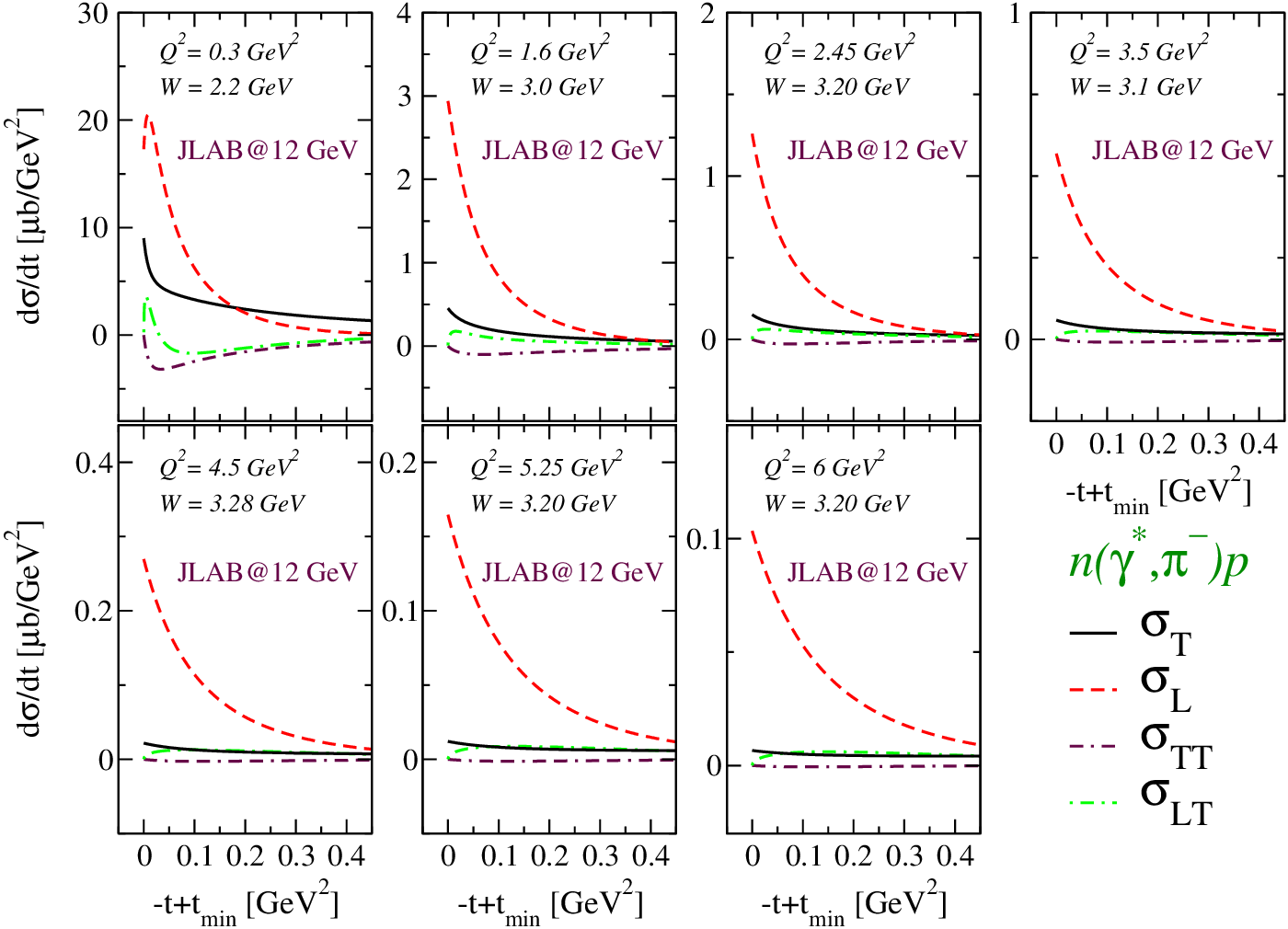}
\caption{\label{JLAB12c} 
(Color online) 
$-t+t_{min}$ dependence of  
\textsc{l/t} separated differential cross sections $d\sigma_{\rm T}/dt$
(solid), $d\sigma_{\rm L}/dt$ (dashed), $d\sigma_{\rm LT}/dt$ (dash-dotted)
and  $d\sigma_{\rm TT}/dt$ (dash-dash-dotted) in exclusive reaction
$n(\gamma^*,\pi^-)p$ in the $(Q^2,W)$ kinematics at JLAB@12.
\vspace{-0.2cm}
}
\end{figure*}

\subsection{HERMES data}
In Section~\ref{DIS_Hermes} we concluded that the physics content of the HERMES
deep exclusive $p(\gamma^*,\pi^+)n$ data is essentially the same as at JLAB.
Figure~\ref{HermesQ2dep} shows the $Q^2$ dependence of 
the measured cross sections in DIS for different 
$x_{\rm B}$ bins~\cite{:2007an,AirapetianPriv}. 
These are the same data sets from HERMES (see previous section) integrated over $-t$. 
The $Q^2$ dependence of the 
experimental data is well described by the calculations (solid curves) from 
the present model. The dashed and dash-dotted curve describe the longitudinal
$\varepsilon \sigma_{\rm L}$ and the transverse $\sigma_{\rm T}$ components, respectively.

In the region of small Bjorken $x_{\rm B}$, see left panel in
Figure~\ref{HermesQ2dep}, the integrated longitudinal component dominates over
the transverse cross section. With increasing $x_{\rm B}$ the strength of the
transverse component is increasing and for values of $x_{B}$ in the third bin
the transverse cross section $\sigma_{\rm T}$ becomes the dominant part of the 
exclusive cross section. An increase of the relative contribution of 
$\sigma_{\rm T}$ as a function of $x_{\rm B}$ can be clearly seen in the 
right panel of Figure~\ref{HermesQ2dep} where $0.26<x_{\rm B}<0.55$.
There the first $Q^2$ bin corresponds to the average value of $x_{\rm B}=0.29$ and
the last $Q^2$ bin to the average value of $x_{\rm B}=0.44$~\cite{AirapetianPriv}.

\section{Beam spin asymmetry}
In this work we have restricted ourselves to exclusive 
$(e,e'\pi^{\pm})$ reactions with an unpolarized target. With a polarized beam 
$(\vec{e},e'\pi^{\pm})$ and with an unpolarized target there is an additional 
component $\sigma_{\rm LT'}$, Eq.~(\ref{dsdte}), which is
proportional to the imaginary part of an interference between the \textsc{l/t}
photons and therefore sensitive to the relative phases of amplitudes.

In general, a nonzero $\sigma_{\rm LT'}$ or the corresponding 
beam SSA $A_{\rm LU}(\phi)$, Eq.~(\ref{BSSA}), demands 
interference between single helicity flip and nonflip or double helicity flip 
amplitudes. In Regge models the asymmetry may result from 
Regge cut corrections to single reggeon exchange~\cite{Ahmad:2008hp}. 
This way the amplitudes in the product acquire different phases and therefore 
relative imaginary parts. A nonzero beam SSA  can be also generated by the 
interference pattern of amplitudes where particles with opposite
parities are exchanged.
 
In the following we discuss briefly the generic features of the beam SSA in the present
model. The comparative analysis of the SSA at JLAB and HERMES will
be presented in the forthcoming publication.

In the left panel of Figure~(\ref{BSAAvakian}) we plot the CLAS 
data~\cite{Avakian:2004dt} for the azimuthal moment $A^{\sin(\phi)}_{\rm LU}$ 
associated with the beam SSA, Eq.~(\ref{BeamSSAmoment}), in the reaction 
$p(\vec{e},e'\pi^+)n$.  These data have been collected in hard scattering
kinematics $E_e=5.77$~GeV, $W>2$~GeV and $Q^2>1.5$~GeV$^2$. 
The experiment shows a sizable and positive beam SSA. 
In the left and right panels of Figure~(\ref{BSAAvakian}) 
we present the results for the azimuthal moments $A^{\sin(\phi)}_{\rm LU}$ 
in the reactions $p(\vec{e},e'\pi^+)n$ and $n(\vec{e},e'\pi^-)p$,
respectively. 
The $(Q^2,W)$ binning of the experimental data point is not available. 
In the following the calculations are done for the lowest $(Q^2,W)$ bin 
corresponding to the values of $Q^2=1.5$~GeV$^2$ and $W^2=4$~GeV$^2$.

At first, we consider $A^{\sin(\phi)}_{\rm LU}$ generated by the exchange of 
Regge trajectories. In Figure~(\ref{BSAAvakian}) the dashed curves describe  
the model results without the \rp effects and neglecting the exchange of the
axial-vector $a_1(1260)$ Regge trajectory. This model results in a zero
$A^{\sin(\phi)}_{\rm LU}$ and therefore a zero beam SSA. The addition of the unnatural
parity $a_1(1260)$-exchange generates by the interference with the natural parity 
$\rho(770)$ exchange a sizable $A^{\sin(\phi)}_{\rm LU}$ in both channels. This result
corresponds to the dash-dotted curves in Figure~(\ref{BSAAvakian}). 
In the rest of observables discussed above the effect of the axial-vector 
$a_1(1260)$ is small. However, as one can see, the contribution of $a_1(1260)$ 
is important in the polarization observables. For instance, a strong
interference pattern of the $a_1(1260)$-reggeon exchange makes the
polarization  observables, like the beam SSA,  very sensitive to the different 
scenarios~\cite{Bechler:2009me} describing the structure and behavior 
of $a_1(1260)$ in high-$Q^2$ processes.

In the last step we account for the \rp contribution. The latter
strongly influence the asymmetry parameter
$A^{\sin(\phi)}_{\rm LU}$. The model results (solid curves) are in agreement 
with the positive $A^{\sin(\phi)}_{\rm LU}$ in the
$\pi^+$ channel and predict much smaller $A^{\sin(\phi)}_{\rm LU}$ in the $\pi^-$
channel.  A sizable and positive $A^{\sin(\phi)}_{\rm LU}$ has been also observed at HERMES
in $\pi^+$ SIDIS close to the exclusive limit $z\to
1$~\cite{Airapetian:2006rx}.

\section{A benchmark for JLAB at 12 GeV: $p(\gamma^*,\pi^+)n$ at Cornell}

A forthcoming upgrade of the JLAB to 12 GeV will allow to measure 
$p(e,e'\pi^+)n$ and  $n(e,e'\pi^-)p$ reactions for values of 
$Q^2=1.6 \div 6.0$~GeV$^2$ and $W$ near 3~GeV~\cite{HuberJLAB12prop}. 
This is just an intermediate region between the present JLAB and the deep
exclusive HERMES data. In this $(Q^2,W)$ region there are old Cornell 
data~\cite{Cornell_2} around $W\simeq 3.1$~GeV and values of 
$Q^2\simeq 1.2$ and 1.7~GeV$^2$. These data may serve as a benchmark 
for the JLAB at 12 GeV predictions.

Figure~\ref{Cornell3GeV12and17} shows the Cornell data~\cite{Cornell_2} and
the calculated differential cross sections for the electroproduction of
$\pi^+$ meson as a function of $-t$. As an example, we selected data with the 
virtual-photoproduction planes of the emitted pions
in average parallel $-45^{\circ} < \phi < 45^{\circ}$ 
to the electron scattering plane. 
The solid curves in Figure~\ref{Cornell3GeV12and17} describe the model results and
the dashed curves correspond to the results without the resonance
contributions. The cross sections have been integrated over the corresponding
range of azimuthal out-of-plane angles.
Figure~\ref{CornellRatio3GeV12} shows the calculated ratio of $\pi^-$ and $\pi^+$
differential cross sections (solid curve) as a function of $-t$. The data are from
\cite{Cornell_2}. As in Figure~\ref{Cornell3GeV12and17}
the cross sections have been integrated over
the range of azimuthal out-of-plane acceptance $-45^{\circ} < \phi < 45^{\circ}$.

The difference between the solid and dashed curves in 
Figure~\ref{Cornell3GeV12and17} comes from the contribution of
resonances. The latter effect is expected to be
important at JLAB@12, see Section~\ref{JLAB12} for the results.

\section{ JLAB at 12 GeV}
\label{JLAB12}
Finally, in this section we provide the predictions for the  \textsc{l/t} separated
$\pi^{\pm}$ differential cross sections 
at forward angles in the $(Q^2,W)$ kinematics proposed for the forthcoming 
$F\pi$-12 experiment~\cite{HuberJLAB12prop} at JLAB. 
The primary goal of the measurements is an
extraction of the pion form factor from the longitudinal data at high values of $Q^2$.

In Figures~\ref{JLAB12b} and~\ref{JLAB12c} 
we plot the $-t+t_{min}$ dependence of the \textsc{l/t} partial differential
cross sections  $d\sigma_{\rm T}/dt$
(solid), $d\sigma_{\rm L}/dt$ (dashed), $d\sigma_{\rm LT}/dt$ (dash-dotted)
and  $d\sigma_{\rm TT}/dt$ (dash-dash-dotted)
in the reactions $p(\gamma^*,\pi^+)n$ and $n(\gamma^*,\pi^-)p$, respectively. 
In these calculations we used the value of $\Lambda_{\gamma\pi\pi}^2=0.46$~GeV$^2$.

As one can see, at JLAB@12 in exclusive reaction $p(\gamma^+,\pi^+)n$ 
the transverse cross section $d\sigma_{\rm  T}/dt$ gets smaller compared to JLAB@5, see solid curves 
in Figures~\ref{JLAB12b}.
But $d\sigma_{\rm  T}/dt$ still gives important contributions at
forward angles. The ratios $R=d\sigma_{\rm T}/d\sigma_{\rm L}$ of the transverse and 
longitudinal cross sections at forward $\pi^+$ angles $t=t_{min}$ for $W$ reached
at JLAB@5 and JLAB@12 are compared in Table~\ref{RatioTableJLAB12}. For the
comparison the values of $Q^2=1.6$ and 2.45~GeV$^2$ are used. 
With increasing value of $W$ at fixed $Q^2$ the ratio gets smaller and makes an accurate 
determination of the longitudinal cross section needed for the extraction 
of the pion from factor feasible. 

In the $\pi^-$ channel the contribution of the tranverse and interference
cross sections is predicted to be much smaller. As one can see in
Figure~\ref{JLAB12c} the $\pi^-$ electroproduction cross section is 
largely longitudinal (dashed curves). If true this may provide a complimentary 
and probably more reliable access to the pion form factor from exclusive 
$\pi^-$ electroproduction off the deuteron target.

\begin{table}[b]
\begin{tabular}{c|ccc}
JLAB & $Q^2$ & $W$ & $R=d\sigma_{\rm T}/d\sigma_{\rm L}$ \\ 
     &  [GeV$^2$] & [GeV] &  \\
\hline
5 GeV  & 1.60 & 2.2 & 0.43\\
       & 2.45 & 2.2 & 0.68\\
\hline 
12 GeV & 1.60 & 3.0 & 0.28\\
       & 2.45 & 3.0 & 0.32 \\
\hline 
\end{tabular}
\caption{\label{RatioTableJLAB12}The ratio $R=d\sigma_{\rm T}/d\sigma_{\rm L}$ 
of the transverse and longitudinal cross sections at forward angles
$t=t_{min}$ in a kinematics of JLAB@5 with $W=2.2$~GeV and JLAB@12 with
$W=3$~GeV  and values of $Q^2=1.6$ and 2.45~GeV$^2$.}
\end{table}

\section{Summary}
In summary, a description of exclusive charged pion electroproduction 
$(e,e'\pi^{\pm})$ off nucleons at high energies is proposed.
Following a two-component hadron-parton picture of Refs.~\cite{Kaskulov:2008xc,Kaskulov:2009gp}
the  model combines a Regge pole approach with residual effects of nucleon
resonances. The contribution of nucleon
resonances has been assumed to be dual to direct partonic
interaction and therefore describes the hard part of the model cross
sections. The resonance/partonic effects are taken into account using a  
Bloom-Gilman connection between the exclusive hadronic form factors
and inclusive deep inelastic structure functions. 
In the soft hadronic sector the exchanges of $\pi$(140), vector $\rho(770)$
and axial-vector $a_1(1260)$ and $b_1(1235)$ Regge trajectories have been considered.  

We have shown that with only a few physical assumptions a quantitative description of
exclusive $\pi^{+}$ and $\pi^-$  electroproduction data can be achieved in a large range of $(Q^2,W)$ 
from JLAB to DIS region at HERMES.
In particular, the \textsc{l/t}
partial longitudinal, transverse and interference cross sections measured at JLAB
and DESY are reproduced. Our principal result 
is that a longstanding problem concerning the description of the transverse and
interference cross sections can be solved by the resonance/partonic effects
in line of the DIS mechanism proposed 
in~\cite{Kaskulov:2008xc,Kaskulov:2009gp}. 
However, the present model goes beyond the two-component approach 
of~\cite{Kaskulov:2008xc,Kaskulov:2009gp} and allows to treat the resonance/partonic
contributions on the amplitude level.
The latter show up
as a large transverse background contribution
to the $\pi$ quasi-elastic knockout mechanism. 
As in~\cite{Kaskulov:2008xc,Kaskulov:2009gp}, we find that at high values of
$Q^2$ the resonances dominate in $\sigma_{\rm T}$. 

The contribution of resonances  in the forward longitudinal response 
$\sigma_{\rm L}$ is rather small and makes an experimental 
isolation of the pion-pole amplitude and the pion transition form factor in
the region of Bjorken $x_{\rm B} < 0.5$ feasible.
The interference pattern of the $\pi$-exchange and resonance
contributions is sufficient to explain the sign and magnitude of the 
interference $\sigma_{\rm TT}$ and $\sigma_{\rm LT}$ cross sections 
measured at JLAB and DESY. The same resonance/partonic mechanism
is responsible for the positive azimuthal beam  SSA  observed in $p(\vec{e},e'\pi^+)n$. 
On the contrary, the beam SSA in deep exclusive $\pi^-$ production off the neutrons
is predicted to be much smaller in magnitude and very sensitive to
the different scenarios concerning the structure of the $a_1(1260)$ meson.

The $Q^2$ behavior of the model exclusive $p(\gamma^*,\pi^+)n$ reaction cross sections 
is in agreement with JLAB and deeply virtual HERMES data. 

We have furthermore calculated the ratio of $\pi^-/\pi^+$ cross sections which is of present
interest in the dedicated experiments at JLAB. 
Model predictions for JLAB at 12 GeV are also provided. 
On the experimental side, the present results may be used as a
guideline in the experimental  analysis of background contributions
to the $\pi$ quasi-elastic knockout mechanism. The latter is important for the 
extraction of the pion transition form factor to 
minimize systematic uncertainties.

\begin{acknowledgments}
This work was supported by DFG through the SFB/TR16.
\end{acknowledgments}

\begin{appendix}
\section{\label{appSCS} The partial virtual-photon nucleon cross sections}
The longitudinal/transverse (\textsc{l/t}) separated photon-nucleon cross sections
take the forms
\be
\label{LTcsL}
\frac{1}{\mathcal{N}}
\frac{d\sigma_{\rm L}}{dt} =
\overline{(J^{\mu}\epsilon_{\mu}^{0} J^{\nu\dagger}\epsilon_{\nu}^{\dagger 0})},
\ee

\be
\label{LTcsT}
\frac{1}{\mathcal{N}}\frac{d\sigma_{\rm T}}{dt} =
\frac{1}{2}\sum \limits_{\lambda=\pm 1}
\overline{(J^{\mu}\epsilon_{\mu}^{\lambda} J^{\nu\dagger}\epsilon_{\nu}^{\dagger\lambda})},
\ee

\be
\label{LTcsTT}
\frac{1}{\mathcal{N}}\frac{d\sigma_{\rm TT}}{dt} =
-\frac{1}{2}\sum \limits_{\lambda=\pm 1}
\overline{(J^{\mu}\epsilon_{\mu}^{\lambda} J^{\nu\dagger}\epsilon_{\nu}^{\dagger-\lambda})},
\ee

\be
\label{LTcsLT}
\frac{1}{\mathcal{N}}\frac{d\sigma_{\rm LT}}{dt} =
-\frac{1}{2\sqrt{2}}\sum \limits_{\lambda=\pm 1} \lambda
[\overline{(J^{\mu}\epsilon_{\mu}^{0}
  J^{\nu\dagger}\epsilon_{\nu}^{\dagger\lambda})}
+ \overline{(J^{\mu}\epsilon_{\mu}^{\lambda}
  J^{\nu\dagger}\epsilon_{\nu}^{\dagger 0})
}],
\ee

\be
\label{LTcsLTpr}
\frac{1}{\mathcal{N}}\frac{d\sigma_{\rm LT'}}{dt} =
-\frac{1}{2\sqrt{2}}\sum \limits_{\lambda=\pm 1} \lambda
[\overline{(J^{\mu}\epsilon_{\mu}^{0}
  J^{\nu\dagger}\epsilon_{\nu}^{\dagger\lambda})}
- \overline{(J^{\mu}\epsilon_{\mu}^{\lambda}
  J^{\nu\dagger}\epsilon_{\nu}^{\dagger 0})
}],
\ee
where $\overline{(...)}$ stands for the sum and average over the initial and
final nucleon spins.
The normalization factor reads
\be
\mathcal{N}= \frac{\alpha_e}{4\pi} \frac{2\pi \, M^2_N}{(W^2-M^2_N)W q^{*}},
\ee
where $q^*$ is a three momentum of the incoming virtual photon in the
$\gamma^*N$ center of mass frame.  
$\epsilon_\mu^{\lambda}$ are the basis
vectors of circular polarization for the virtual photon with helicities
$\lambda{=}\pm 1, 0$
quantized along the three momentum $\vec{q}$, {\it i.e.},
\begin{eqnarray}
\label{PhotonHStates1}
\epsilon^{\pm}_{\mu} &=& \mp \frac{1}{\sqrt{2}} (0,1,\pm i,0), \\
\epsilon^{0}_{\mu}   &=& \frac{1}{\sqrt{Q^2}}(\sqrt{\nu^2+Q^2},0,0,\nu),
\end{eqnarray}
and $J^{\mu}$ is the nuclear transition axial-vector
current describing the pion
production in the momentum space.

\section{\label{GaugedBorn} On the gauged electric amplitude}
The  Lorentz tensor-vector decomposition of the nucleon-pole term in
Eq.~(\ref{PipolePl}) is given by
\begin{widetext}
\begin{eqnarray}
\label{NBtransf}
\bar{u}_{s'}(p') \gamma_5 \frac{(p+q)_{\sigma}\gamma^{\sigma}\gamma^{\mu} +
    M_p\gamma^{\mu}}{s-M_p^2+i0^+}u_s(p)
= \bar{u}_{s'}(p')\gamma_5 
\left[
\frac{i\sigma^{\mu\sigma}q_{\sigma}}{s-M_p^2+i0^+} 
+ \frac{(p+p'+ k')^{\mu}}{s-M_p^2+i0^+}
\right] u_s(p),
\end{eqnarray}
\end{widetext}
where $\sigma^{\mu\sigma}=\frac{i}{2}[\gamma^{\mu},\gamma^{\sigma}]$.
The axial-tensor term in the {\it r.h.s.} of Eq.~(\ref{NBtransf}) is 
gauge invariant by itself. It turns out that only the orbital part 
proportional to $p+p'+k'$ is needed to conserve the charge in the sum 
with the $\pi$-pole amplitude, the first term in Eq.~(\ref{PipolePl}).
In photoproduction the orbital part has no physical significance
because it appears in the physical scattering amplitude multiplied by 
the polarization vector $\epsilon_{\mu}^{\lambda}$.  Since, 
$q^{\mu}\epsilon_{\mu}^{\pm}=p^{\mu}\epsilon_{\mu}^{\pm}=0$
the product $(p+p'+k')\epsilon_{\mu}^{\pm} = (2p+q)\epsilon_{\mu}^{\pm}=0$.
The nucleon-pole term generates a large axial-tensor
background to the meson-pole amplitude which has important consequences in
photoproduction. For instance, since the interference of $\pi$ with $\rho(770)$ is
trivially zero it is $\sigma^{\mu\sigma}\gamma_5$ in the nucleon-pole term which,
by interference with $\rho$, is responsible for the $\pi^-/\pi^+$ asymmetry. 
It is also this term which explains the forward peak in the forward
$\pi^{\pm}$ production and the rapid variation of the polarized photon
asymmetry in the same region.

\end{appendix}

\end{document}